\documentclass[final,twocolumn]{elsarticleMod}
\usepackage{amsmath,amssymb,amsfonts}
\usepackage{bbm,bm}
\usepackage[english]{babel}
\usepackage{datetime,dsfont}
\usepackage{enumerate}
\usepackage[letterpaper,top=2cm,bottom=2cm,left=3cm,right=3cm,marginparwidth=1.75cm]{geometry}
\usepackage{graphicx}
\usepackage{lineno}
\usepackage{subcaption,slashed,siunitx}
\usepackage{ulem}
\usepackage[svgnames]{xcolor}
\usepackage{tikz}

\usepackage{hyperref}
\hypersetup{colorlinks=true, linkcolor=blue, citecolor=ForestGreen,
filecolor=magenta, urlcolor=ForestGreen,
  pdftitle={Javurkova, et al,
  Phys. Lett. B 855 (2024), 138787 [2401.17365]},
  pdfauthor={Javurkova, et al},
  pdfsubject={Subject},pdfcreator={Creator},pdfproducer={Producer},
  pdfkeywords={Helicity Polarization}{Standard Model Effective Field Theory}{Large Hadron Collider}{Electroweak Physics}
}
\newcommand*{\round}[1]{\num[output-decimal-marker={.},
                             round-mode=figures,
                             round-precision=3,
                             group-digits=false]{#1}}

\newcommand{\mgamc}{\texttt{mg5amc}}
\newcommand{\mgFull}{\texttt{MadGraph5\_aMC@NLO}}
\newcommand{\libName}{\texttt{SM\_Loop\_VPolar}}

\def\fb{{\rm\ fb}}

\newcommand{\invfb}{{\rm ~fb^{-1}}}

\def\GeV{{\rm\ GeV}}
\def\TeV{{\rm\ TeV}}

\definecolor{magenta}{HTML}{FF00FF}
\definecolor{cornflowerblue}{HTML}{6495ED}
\definecolor{turquoise}{HTML}{40E0D0}

\definecolor{darkgreen}{rgb}{0.0, 0.2, 0.13}
\definecolor{darkmagenta}{rgb}{0.55, 0.0, 0.55}
\definecolor{amber}{rgb}{1.0, 0.6, 0.0}

\newcommand{\orcid}[1]{\,\href{https://orcid.org/#1}{\includegraphics[width=9pt]{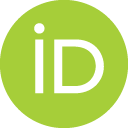}}}
\newcommand{\orcidMJ}{0000-0001-8798-808X} 
\newcommand{\orcidRR}{0000-0002-3316-2175} 
\newcommand{\orcidRCLSA}{0000-0001-5200-9195} 
\newcommand{\orcidJS}{0000-0002-6016-8011} 

\graphicspath{{figures/}}

\journal{Preprint numbers: IFJPAN-IV-2024-2, COMETA-2024-01} 
\begin{document}
\begin{frontmatter}

\title{Polarized $ZZ$ pairs in gluon fusion and vector boson fusion at the LHC}

\author[umass,ubel]{Martina Javurkova\ \orcid{\orcidMJ}} \ead{martina.javurkova@cern.ch (contact author)}
\author[ifj]{Richard Ruiz\ \orcid{\orcidRR}} \ead{rruiz@ifj.edu.pl}
\author[umass]{Rafael Coelho Lopes de S\'a\ \orcid{\orcidRCLSA}}\ead{rclsa@umass.edu}
\author[umass]{Jay Sandesara\ \orcid{\orcidJS}} \ead{jsandesara@umass.edu}

\address[umass]{Physics Department -- University of Massachusetts Amherst, Amherst, Massachusetts, USA }
\address[ubel]{Physics Department -- Matej Bel University, Tajovského 40, Banská Bystrica, 97401, Slovakia}
\address[ifj]{Institute of Nuclear Physics -- Polish Academy of Sciences {\rm (IFJ PAN)}, ul. Radzikowskiego, Krak{\'o}w, 31-342, Poland}


\begin{abstract}
Pair production of helicity-polarized weak bosons $(V_\lambda=W^\pm_\lambda, Z_\lambda)$ from gluon fusion $(gg\to V_\lambda V'_{\lambda'})$ and weak boson fusion $(V_1V_2\to V_\lambda V'_{\lambda'})$ are powerful probes of 
the Standard Model, 
new physics,
and properties of quantum systems.
Measuring cross sections of polarized processes is a chief objective of the Large Hadron Collider's (LHC) Run 3 and high luminosity programs, but progress is limited by the simulation tools that are presently available.
We propose a method for computing polarized cross sections that works by directly modifying Feynman rules instead of (squared) amplitudes.
The method is applicable to loop-induced processes,
and can capture
the interference between arbitrary polarization configurations,
interference with non-resonant diagrams,
as well as off-shell/finite-width effects.
By construction, previous results that work at the
(squared) amplitude level are recoverable.
As a demonstration, we report the prospect 
of observing and studying polarized $Z_\lambda Z_{\lambda'}$ pairs when produced via gluon fusion and electroweak processes in final-states with four charged leptons at the LHC, using the new method to simulate the gluon fusion process.
Our Feynman rules are publicly available as a set of 
\textit{Universal FeynRules Object} libraries called {\texttt{SM\_Loop\_VPolar}}.

\end{abstract}

\begin{keyword}
{\small
Large Hadron Collider, Helicity Polarization, Monte Carlo Tools, Multiboson Systems
}
\end{keyword}

\end{frontmatter}

\section{Introduction}
\label{sec:intro}

Helicity-polarized vector bosons in the Standard Model (SM) carry information about the mechanism of electroweak (EW) symmetry breaking (EWSB) and its role in ensuring unitarity in the model. Measurements of polarized vector states subsequently provide experimental tests of fine, structural cancellations in the SM. They can also be used to probe  extensions of the SM, such as extended gauge theories~\cite{Pati:1974yy,Mohapatra:1974hk,Mohapatra:1974gc,Senjanovic:1975rk,Senjanovic:1978ev}, 
composite Higgs models~\cite{Kaplan:1983fs,Kaplan:1983sm,Georgi:1984af,Dugan:1984hq,Contino:2003ve,Agashe:2004rs,Contino:2006qr,Agashe:2006at,Bellazzini:2014yua,Panico:2015jxa},
generic anomalous couplings~\cite{Hernandez-Juarez:2023dor,Cao:2020npb},
as well as quantum properties of multiboson systems~\cite{Severi:2021cnj,Aguilar-Saavedra:2022wam,Aguilar-Saavedra:2022mpg,Fabbrichesi:2023cev,Ashby-Pickering:2022umy}.
Such sensitivity holds even in the decoupling limits~\cite{Henning:2018kys},
where it is challenging for direct searches. 
Consequently, 
any program aiming to explore EWSB and new physics in the Higgs sector
at the Large Hadron Collider (LHC) and future collider experiments
will involve measuring polarized vector 
bosons~\cite{Green:2016trm,Anders:2018oin,EuropeanStrategyforParticlePhysicsPreparatoryGroup:2019qin,EuropeanStrategyGroup:2020pow,Covarelli:2021gyz,BuarqueFranzosi:2021wrv}.

Despite the physics needs and motivations, 
there is a limited supply of predictions for 
integrated $(\sigma)$  and differential $(d\sigma)$ cross sections 
for TeV-scale proton $(p)$ collisions 
with 
intermediate or final-state helicity-polarized particles, e.g.,
\begin{align}
 pp \to V_\lambda V'_{\lambda'}(V''_{\lambda''})\quad \text{for}\   V\in\{W^\pm,Z,\gamma\}\ .
\end{align}
Here and throughout, $\lambda$ denotes the helicity of $V$ and $q$ is its momentum.
Predictions are notably scarce
beyond tree level in perturbation theory.

Typically, helicity-polarized cross sections are obtained
 by first constructing the matrix elements (or their square) for an unpolarized process,
and then decomposing the numerator of
an intermediate, resonant propagator
$V$
into the outer product of one or more pairs of
polarization vectors (or spinors for fermions).
This is known generically
as the ``truncated-propagator'' method.

With the so-called double-pole approximation (DPA)~\cite{Aeppli:1993cb,Aeppli:1993rs,Denner:2000bj,Billoni:2013aba},
also known  as the on-shell projection (OSP) technique,
non-resonant diagrams are neglected
and momenta $q$ in the numerators of propagators are ``projected'' to be on mass shell.
($q^2$ factors in the denominators are unmodified.)
Dedicated predictions up to next-to-leading order (NLO) in both EW and quantum chromodynamics (QCD) are available~\cite{Denner:2021csi,Le:2022lrp,Denner:2023ehn,Dao:2023kwc,Pelliccioli:2023zpd}
as are predictions up to next-to-next-to-leading order (NNLO) in QCD~\cite{Pellen:2021vpi,Poncelet:2021jmj}. However, these state-of-the-art predictions are limited to at most two intermediate, helicity-polarized weak bosons.
Leading order (LO) event generators employing DPA/OSP are available, but again are mostly restricted to select processes~\cite{Ballestrero:2007xq,Ballestrero:2011pe,Ballestrero:2017bxn,Ballestrero:2019qoy,Hoppe:2023uux}.

By combining discrete Monte Carlo sampling over helicities
with the spin-correlated narrow width approximation (NWA),
a generalization of the truncated-propagator method
was developed~\cite{BuarqueFranzosi:2019boy} that did not require modifying intermediate momenta.
This led to the automation of helicity-polarized cross sections
at LO for arbitrary tree-level processes~\cite{BuarqueFranzosi:2019boy}.
More recently, the computation of polarized cross sections
with the NWA
was automated at (approximately) NLO in QCD for arbitrary processes by projecting unpolarized cross section onto a basis of polarizations, and hence providing polarized cross sections~\cite{Hoppe:2023uux}.

We attempt to advance this program by proposing another method for computing polarized cross sections that extends the truncated-propagator method 
to individual Feynman rules.
The broad idea is that the (coherent/interfering) sum of polarizations of a \textit{single, unpolarized}
intermediate vector boson
can be realized as the (coherent/interfering) sum of \textit{several, polarized} states.
In other words, the graph of a vector boson (or fermion)
can be split
into the sum of several propagating states.
Polarized cross sections that include full off-shell effects,
interference with non-resonant contributions,
polarized $t$-channel contributions,
and loop-induced contributions can then be obtained via diagram selection, which is a common technique in realistic simulations for the LHC.

As a demonstration and using publicly available
Monte Carlo (MC) simulation tools,
we use the new method to
estimate the sensitivity of the LHC experiments
to helicity-polarized $ZZ$ pairs produced via gluon fusion (ggF).
We also estimate the sensitivity to EW processes,
including  weak boson fusion (VBF) and annihilation channels.
We focus on the data set sizes expected by the end of the LHC's Run 3 era of operations
and with the full dataset of the high-luminosity (HL) LHC.

This work continues as follows:
In Sec.~\ref{sec:PolarTheory} we describe our proposed method for computing polarized cross sections and its practical implementation.
In Sec.~\ref{sec:xsec} we present LHC predictions for the production of  $Z_\lambda Z_{\lambda'}$ pairs from gluon fusion.
In Sec.~\ref{sec:PolarMeas} we present our 
projections for measuring polarized $ZZ$ cross sections.
Further applications are discussed in Sec.~\ref{sec:SpinCorr}, and we conclude in Sec.~\ref{sec:Conclusion}.
\ref{app:implementation}-\ref{app:setup_mc}
contain additional technical details to reproduce our work.


\section{Helicity polarization as a Feynman rule}
\label{sec:PolarTheory}

After EWSB, the propagator of a massive, on-shell gauge boson $V$ with momentum $q$ and mass $M_V$ is related to its polarization vectors $(\varepsilon)$ in the Unitary gauge by the completeness relationship
\begin{align}
\label{eq:completeness}
    g_{\mu\nu} - q_\mu q_\nu / M_V^2
    =
    \sum_{\lambda\in\{0,\pm1\}} 
    \varepsilon_\mu(q,\lambda)\ \varepsilon^*_\nu(q,\lambda)\ .
\end{align}
The sum runs over the longitudinal $(\lambda=0)$ and transverse $(\lambda=\pm)$ states of $V$. 
Eq.~\eqref{eq:completeness} can be extended to off-shell $V$ by extending the sum to the so-called ``auxiliary/scalar'' polarization $(\lambda=A)$, 
and to the $t$-channel production by 
including $\lambda=A$ and summing over $(-1)\times\varepsilon_\mu(\lambda=0)\ \varepsilon^*_\nu(\lambda=0)$ instead of $(+1)\times\varepsilon_\mu(\lambda=0)\ \varepsilon^*_\nu(\lambda=0)$~\cite{Halzen:1984mc,Weinberg:1995mt}.
Strictly speaking, the helicity of $V$ in a particular frame is only well-defined, and hence experimentally accessible, 
when $V$ is on shell.
This is because helicity labels an eigenstate in the mass basis and off-shell states are not eigenstates of this basis.
However, in particular circumstances~\cite{Ruiz:2021tdt},
Eq.~\eqref{eq:completeness} is also meaningful for $t$-channel exchanges~\cite{Dawson:1984gx,Kane:1984bb,Kunszt:1987tk}.

The decoherence of a heavy object's helicity when off-shell is evident in its propagator, which can be identified as a \textit{coherent} sum of external
polarization vectors (or spinors for fermions).
That is to say, using Eq.~\eqref{eq:completeness}
one can express the propagator of $V$ (with mass and width $M_V,\Gamma_V$) as
\begin{align}
    &\Pi_{\mu\nu}^V (q) =
    \frac{-i\left(g_{\mu\nu} - q_\mu q_\nu / M_V^2\right)}{q^2-M_V^2 +iM_V\Gamma_V}\
    \label{eq:full_prop}
    \\
    &=\sum_{\lambda\in\{0,\pm1,A\}} 
    \eta_\lambda\ \left(
    \frac{-i\varepsilon_\mu(q,\lambda)\ \varepsilon^*_\nu(q,\lambda)}{q^2-M_V^2 +iM_V\Gamma_V} \right)\ .
    \label{eq:full_prop_sum}
\end{align}
Here, $\eta_\lambda=+1$, unless $\lambda=0$ and $V_{\lambda}$ is in the $t$-channel; in that case $\eta_\lambda=-1$.
In Ref.~\cite{BuarqueFranzosi:2019boy}, the quantity in the summation of Eq.~\eqref{eq:full_prop_sum} is labeled the ``helicity-truncated'' propagator $\Pi_{\mu\nu}^{V\lambda}$,
\begin{align}
    \Pi_{\mu\nu}^{V\lambda}(q) &= \frac{-i\varepsilon_\mu(q,\lambda)\ \varepsilon^*_\nu(q,\lambda)}{q^2-M_V^2 +iM_V\Gamma_V}\ .
    \label{eq:trunc_prop}
\end{align}

Helicity-polarized matrix elements, 
and hence helicity-polarized cross sections $(d\sigma_\lambda)$, 
can then be defined by using the truncated propagator of Eq.~\eqref{eq:trunc_prop}.
The unpolarized $(\mathcal{M})$
and  polarized  $(\mathcal{M}_\lambda)$ matrix elements
are related by summing over $\lambda$:
\begin{align}
    \mathcal{M}\ &=\ 
    \mathcal{M}_f^\mu\ \left(\sum_{\lambda\in\{\pm1,0,A\}}\ \eta_\lambda\ \times\ \Pi_{\mu\nu}^{V\lambda}\right)\
    \mathcal{M}_i^\nu
\label{eq:me_full}    
    \\
    &=\ \sum_{\lambda\in\{\pm1,0,A\}}\ 
    \eta_\lambda\ \times\
\underset{= \mathcal{M}_\lambda}{\underbrace{\mathcal{M}_f^\mu\cdot \Pi_{\mu\nu}^{V\lambda}\cdot   \mathcal{M}_i^\nu}}\ .
\label{eq:me_decomposition}
\end{align}
Here, $\mathcal{M}_i^\mu$ and $\mathcal{M}_f^\nu$ are the Green's functions (diagram fragments) that sandwich the full propagator $V$, i.e., Eq.~\eqref{eq:full_prop}, in the  unpolarized matrix element.

After squaring the polarized matrix element
and integrating over phase space one obtains $d\sigma_\lambda$. 
The difference between the unpolarized cross section $(d\sigma_{\rm unpol.})$ and summing over all polarized cross sections is the interference between helicity states,
\begin{align}
\label{eq:cross_section}
    d\sigma_{\rm unpol.} &= \nonumber\\ 
    &\sum_{\lambda\in\{\pm1,0,A\}}\ d\sigma_\lambda + 
    \sum_{\lambda,\lambda'}\
    \underset{\mathcal{O}\left(\Re\left[\mathcal{M}^*_\lambda\mathcal{M}_{\lambda'}\right ]\right)}{\underbrace{\text{interference}}}.
\end{align}
The ``interference'' term here also includes the contribution from non-resonant diagrams.

The magnitude of the interference is process dependent but can be incorporated by adjusting the definition of $\Pi_{\mu\nu}^{V\lambda}$ in Eq.~\eqref{eq:trunc_prop} to include multiple polarizations~\cite{BuarqueFranzosi:2019boy}.
For example: one can define the ``transverse''  propagator $\Pi_{\mu\nu}^{VT}$  as the sum over the $\lambda=\pm1$ states.
The interference is also phase-space dependent,
the impact of which can be adjusted by
restrictions on external and intermediate momenta.
Finally, the accuracy of $d\sigma_\lambda$ depends on how many
 intermediate states are polarized.

The traditional truncated-propagator method has several advantages:
(i) off-shell effects are included,
(ii) it is generalizable to fermions and high-spin states, 
(iii) interference between polarizations can be included,
and
(iv) it is applicable to tree- and loop-level processes.
A significant disadvantage, however, 
is the need to modify the propagator routines (software libraries) in existing event generators.
Such work is technically challenging to implement
into event generators capable of simulating arbitrary processes 
because the mappings between particles, 
their Feynman graphs, 
and their propagators are typically hard coded.
Alterations can lead to undesired consequences for unpolarized subgraphs~\cite{BuarqueFranzosi:2019boy,Hoppe:2023uux}.

To ameliorate this difficulty, we propose treating the truncated propagator in Eq.~\eqref{eq:trunc_prop} as a Feynman rule itself.
If one considers the decomposition of a state $V$ as a combination of polarized states $V_\lambda$,
\begin{align}
    \vert V(q,M_V)\rangle\ =\ \sum_{\lambda\in\{\pm1,0,A\}}\ \vert V_{\lambda}(q,M_V,\lambda)\rangle\ ,
\end{align}
and propagates this throughout a Lagrangian,
then graphically one can make the identification:
\begin{align}
\includegraphics[width=.8\columnwidth]{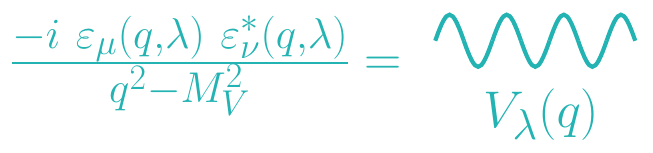}\ .
\label{eq:feynman_rule}
\end{align}
That is, 
the truncated propagator $\Pi_{\mu\nu}^{V\lambda}$ 
in Eq.~\eqref{eq:trunc_prop} 
is the propagator of a vector that has a fixed polarization.
One can then interpret Eq.~\eqref{eq:me_full} as the amplitude for a 
process that is mediated by 
a single particle $V$ with interfering polarizations $\lambda$
and interpret Eq.~\eqref{eq:me_decomposition} as 
the sum of interfering diagrams
for a process mediated by \textit{a collection of particles} $V_\lambda$, where each $V_\lambda$ has its own propagator.
This identification is one chief result of our work.

The key to realizing this approach in realistic simulations 
is the fact that modern, multi-purpose  event generators are sufficiently flexible to handle new Feynman rules\footnote{This is often through an interface to cross-platform  
\texttt{Universal FeynRules Object} (UFO) libraries~\cite{Christensen:2008py,Christensen:2009jx,Degrande:2011ua,Alloul:2013bka}.}, including alternative propagators, as \textit{input} libraries.
Passing Eq.~\eqref{eq:feynman_rule} or Eq.~\eqref{eq:trunc_prop} to event generators as an input library is an efficient alternative to 
modifying the built-in propagators for SM particles, 
which are typically hard-coded.
Moreover, this method is compatible with the automated computation
of so-called $R_2$ and ultraviolet counter-terms at NLO in QCD~\cite{Degrande:2014vpa}.
Therefore,  ``polarization as a Feynman rule''
is also applicable to automated loop-induced processes. 

As a brief comment,
treating polarization as a Feynman rule raises the prospect that  new rules are not
individually Lorentz invariant, and that all external-particle momenta must be written in the same frame through the whole flow of a numerical calculation.
In general, individual Feynman rules are inherently not Lorentz invariant: many rules  in the SM carry spinor or vector indices. As a consequence, numerical implementations of the helicity amplitude method for computing matrix elements (for example: \texttt{HELAS}~\cite{Murayama:1992gi} as implemented in \texttt{MadGraph}~\cite{Stelzer:1994ta}) already require a reference frame to be chosen before phase space integration. For LO predictions in {\mgamc}, including those that are loop-induced, helicity amplitudes are evaluated by default in the partonic frame, though other choices of frame are possible~\cite{BuarqueFranzosi:2019boy}. Our method does not change this operation; rather, the method relies and exploits these generator-dependent capabilities (and their numerical efficiencies).

\subsection{Practical Implementation}
\label{sec:PolarTheory_implement}

To implement our proposal,
we start from the default implementation
of the full SM Lagrangian
into \texttt{FeynRules}~\cite{Christensen:2008py,Christensen:2009jx,Degrande:2011ua,Alloul:2013bka}.
We add four new ``particles'' to the model:   $V_T,\ V_0,\ V_A$ and $V_X$ for $V=W^\pm$ and $Z$, which carry the same properties as the SM $W^\pm$ and $Z$ bosons. We then make a field redefinition $V=V_T+V_0+V_A+V_X$.
The $V_\lambda$ inherit all interactions of the $W$ and $Z$; redundant operators, e.g., $\Delta\mathcal{L}=M_Z^2 Z_T^\mu Z_{A\mu}$, are removed.

We extract QCD renormalization and $R_2$ counter terms (CTs) up to the first order in $\alpha_s$ using
\texttt{FeynRules} {(v2.3.36)} with
\texttt{NLOCT} {(1.02)}~\cite{Degrande:2014vpa}
and \texttt{FeynArts} {(v3.11)}~\cite{Hahn:2000kx}.
At this point, the $V_\lambda$
still have the same propagators as $V$
and subsequently inherit all the CTs associated with $V$
due to the $V=V_T+V_0+V_A+V_X$ redefinitions.
Importantly, the CTs here are purely QCD CTs at $\mathcal{O}(\alpha_s)$, not mixed QCD$\otimes$EW CTs.
This means that our CTs
do not depend on the helicity of $V$.
While the $gg\to e^+e^-\mu^+\mu^-$ process at LO contains no UV  divergencies, $R_2$ terms are present.

We then package all Feynman rules
into a single set of software libraries\footnote{Our public
\texttt{FeynRules} UFO entitled {\libName}
is available from
\href{https://feynrules.irmp.ucl.ac.be/wiki/VPolarization}{feynrules.irmp.ucl.ac.be/wiki/VPolarization}}.
It is at this point that we modify
the propagator library
 so that the ``new'' particles have propagators given by
Eq.~\eqref{eq:trunc_prop}.
Explicitly, $V_T$ carries summed $\lambda=\pm1$ polarizations
and $V_0\ (V_A)$ carries the $\lambda=0\ (A)$ polarization.
$W_X$ and $Z_X$, which are not used in the study,
are copies of the unpolarized SM $W$ and $Z$ that we included for
diagnostic purposes / closure tests.

We use the new UFO ({\libName}) in conjunction with {\mgFull} (\mgamc)
(v3.4.0)~\cite{Stelzer:1994ta,Alwall:2014hca,Hirschi:2015iia}
to build the polarized matrix element,
and hence compute the polarized cross section,
for the full,
loop-induced $2\to4$ process $gg\to e^+e^-\mu^+\mu^-$ at LO,
i.e., at $\mathcal{O}(\alpha_s\alpha^2)$.
We include the full dependence of the top quark~\cite{Hirschi:2015iia}. We do not impose the NWA but remove all diagrams with  photons; we justify removing photons in the next section.
This level of diagram removal is a built-in 
ability {\mgamc}~\cite{Alwall:2014hca,Hirschi:2015iia}.
For unpolarized propagators, we obtain
16 diagrams mediated by top triangles and top boxes, including diagrams with an internal Higgs,
and obtain 14 diagrams each for
massless $u$- and $d$-quark loops.

According to our method, all intermediate $Z$ subgraphs, including non-resonant subgraphs, are split into ``polarized'' subgraphs. 
(For multiboson processes, this can generate a large diagram multiplicity.)
For the doubly polarized matrix elements
$(\lambda_1,\lambda_2)=(0,0)$ and $(T,T)$,
we remove all diagrams containing undesired $V_\lambda$.
Again, this type of diagram removal is a built-in option in {\mgamc}.
In both cases, we obtain 16 (14) diagrams when the top quark (a light quark) runs in the loop.

For the mixed polarized channel $(\lambda_1,\lambda_2)=(0,T)+(T,0)$, we exploit the topology, coupling order, absence of photons.
These conditions dictate that two exactly
$Z_\lambda$ propagators must appear in the full $2\to6$ process.
We employ a type diagram filtering accept/reject diagrams\footnote{\label{foot:gitlab}Removing loop-level diagrams in {\mgamc} is a highly nontrivial issue.
Our routine as implemented in  {\mgamc} is available from the 
repository \href{https://gitlab.cern.ch/riruiz/public-projects/-/tree/master/VPolar_ggZZ}{gitlab.cern.ch/riruiz/public-projects/-/tree/master/VPolar\_ggZZ}.}.
For diagrams induced by triangle loops,
we require that exactly one $Z_0$ and one $Z_T$ appear in
any multi-vertex, tree subgraph attached to the loop.
For diagrams induced by boxes,
we require that one of the tree subgraphs attached to the loop contains $Z_0$ and another contains $Z_T$.
We generate the expected 32 (28) diagrams for top (light) quark loops.

The matrix element for our closure test is obtained by removing
all diagrams with photons and the auxiliary ``$Z_X$'' state.
We otherwise keep all diagrams with $Z_\lambda\in\{Z_T,Z_0,Z_A\}$.
For top quark diagrams,
this generates $16\times3^2=144$ individual diagrams
while $14\times3^2=126$ diagrams are generated for loops with light quarks. 

Investigating polarization rates in 
loop-induced triple or quadruple boson production 
is outside the scope of this work.
However, we have checked that the correct number of diagrams
are generated for
$gg\to Z_{\lambda_1}Z_{\lambda_2}Z_{\lambda_3}\to
e^+e^-\mu^+\mu^-\tau^+\tau^-$.
For the unpolarized case and assuming a massless tau, 
we obtain 228 (186) resonant and non-resonant 
diagrams for top (light) quark loops,
and which includes 6 diagrams with four intermediate $Z$ bosons.
We obtain the same number of diagrams for the
triple polarization configuration
$(\lambda_1,\lambda_2,\lambda_3)=(0,0,0)$.
When all polarizations are allowed, $\lambda_k\in\{0,T,A\}$,
we obtain $n_{\rm top}=(228-6)\times3^3 + (6)\times3^4=6480$ diagrams with a top quark loop, and $n_{\rm light}=(186-6)\times3^3 + (6)\times3^4=5346$ diagrams for each light quark in the loop.



\section[\texorpdfstring{Polarized $Z_\lambda Z_{\lambda'}$ from gluon fusion}{Polarized ZZ from gluon fusion}]
{Polarized $Z_\lambda Z_{\lambda'}$ from gluon fusion}
\label{sec:xsec}

As a first demonstration of our method we present 
total and differential
predictions for polarized $Z_\lambda Z_{\lambda'}$ production from ggF.
We construct our matrix elements and
simulate our signal processes at LO with
{\mgamc} as described in Sec.~\ref{sec:PolarTheory_implement}.
We use the NNPDF31+LUXqed NLO parton distribution function (PDF) (\texttt{lhaid=324900})~\cite{Manohar:2016nzj,Manohar:2017eqh,Bertone:2017bme}. 
DGLAP evolution and PDF uncertainty extraction are
handled by \texttt{LHAPDF} {(v6.3.0)}~\cite{Buckley:2014ana}.
Further simulation details are given in \ref{app:setup_mc}.

\subsection{Polarized cross sections in proton collisions}
\label{sec:xsec_fiducial}

\begin{figure}[!t]
\centering
\includegraphics[width=\columnwidth]{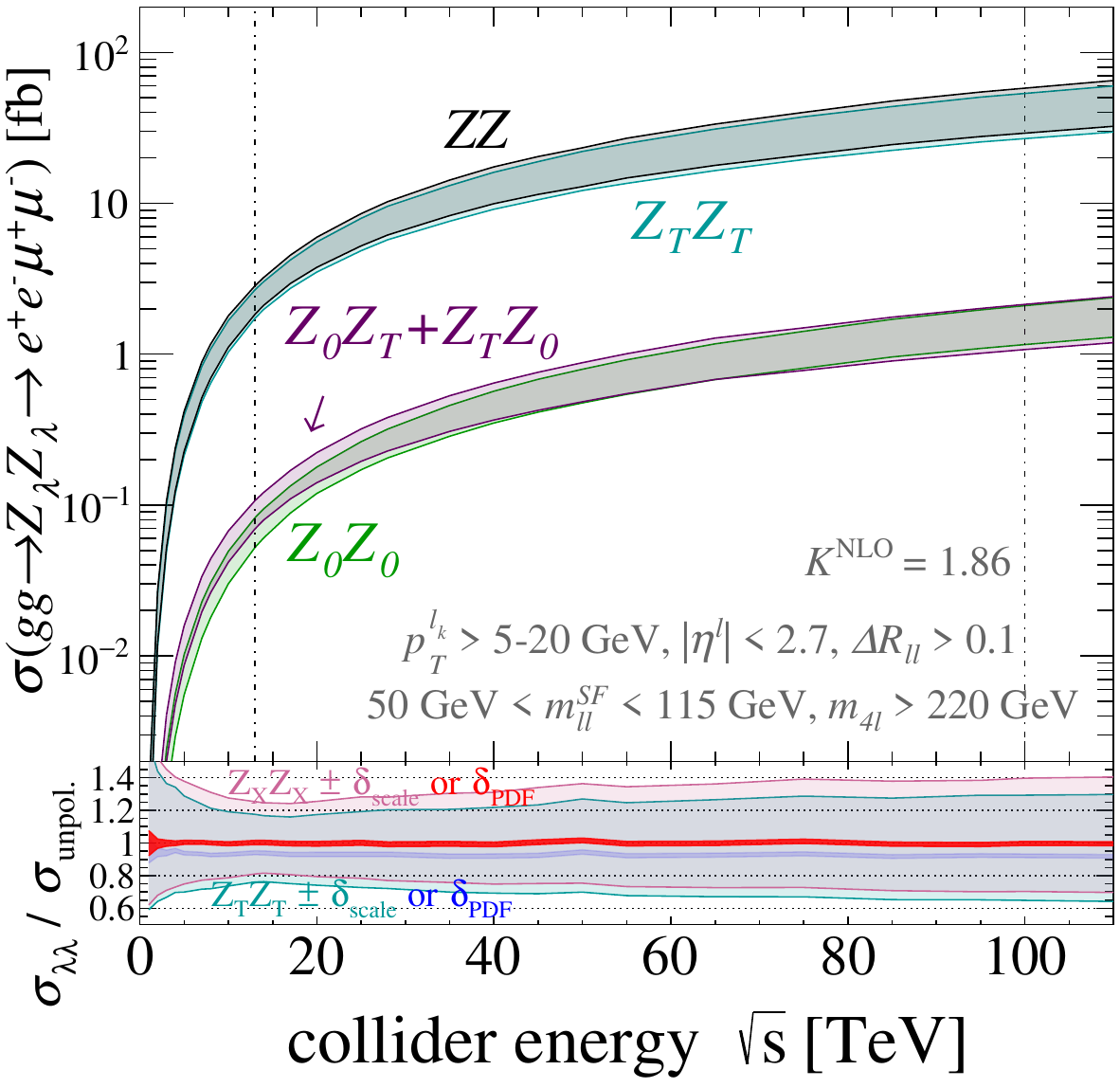}
	\caption{Upper panel:
	As a function of $\sqrt{s}$, the unpolarized $gg\to ZZ\to e^+e^-\mu^+\mu^-$
	cross section (black band) with phase space cuts of Eq.~\eqref{eq:cuts},
	along with the analogous
	$Z_TZ_T$ (blue),
	$Z_TZ_0$ (purple), and
	$Z_0Z_0$ (green) rates.
	Band thickness denotes the scale uncertainty.
	Lower panel: The ratio of the $Z_TZ_T$ (blue) and $Z_XZ_X$ (red) cross sections and their uncertainties relative to the unpolarized rate.
    ($Z_X$ is the sum of the $Z_T$, $Z_0$, and $Z_A$ channels at the amplitude level.)
    Larger (transparent) bands are the scale uncertainties;
    smaller (solid) bands are the PDF uncertainties.}
 \label{fig:xsec}
\end{figure}

In Fig.~\ref{fig:xsec} (top panel) we show
the hadron-level cross section [fb]
as a function of collider energy $\sqrt{s}$ [TeV]
for the full loop-induced, $2\to4$ process
\begin{align}
\label{eq:proc_gf}
    gg\ \to\ Z_{\lambda}Z_{\lambda'}\ \to\ e^+e^-\mu^+\mu^-
\end{align}
at LO for
unpolarized $ZZ$ pairs (black band)
as well as polarized
$Z_TZ_T$ (blue band),
$Z_TZ_0$ (purple band),
and
$Z_0Z_0$ (green band) pairs.
The $Z_TZ_0$ rate includes both $(T,0)$ and $(0,T)$ helicity configurations.
Helicity is defined in the $(gg)$ frame.
The unpolarized prediction does not employ the methodology of Sec.~\ref{sec:PolarTheory} and serves as a check.
Band thickness corresponds to the scale uncertainty.

We remove photon diagrams, 
which are less important around the $Z$ resonance, 
to make the relative importance of individual polarization configurations clearer.
We also impose the following phase space cuts on the  charged leptons:
\begin{subequations}
\label{eq:cuts}
    \begin{align}
        p_T^{\ell_{1~(2)~[3]~\{4\}}}
        & >20~(15)~[10]~\{5\}\GeV,
        \\
        m_{4\ell} > 220\GeV,
        &\
        \vert\eta^\ell\vert < 2.7,\ \Delta R_{\ell\ell} > 0.1
        \\
        50\GeV & < m_{\ell\ell}^{\rm SF} < 115\GeV\ , 
    \end{align}
\end{subequations}
where SF denotes same-flavor lepton pairs
and charged leptons are ranked by $p_T$, i.e., $p_T^i > p_T^{i+1}$.
These generator cuts reflect selection cuts that would be imposed in a real analysis in ATLAS or CMS. 
For all polarization channels and collider energies, we 
estimate\footnote{Ref.~\cite{Alioli:2021wpn} 
reports an NLO $K$-factor of $K^{\rm NLO}\approx1.53$
when using a (N)LO PDF with $\alpha_s(M_Z)\approx0.130~(0.118)$ for the (N)LO computation. To account for this, we reweight $K^{\rm NLO}$ by 
$(\alpha^{\rm 1-loop}_s / \alpha^{\rm 2-loop}_s)^2\approx 1.21$.
We neglect the additional impact due to the different choices of $\mu_f,\mu_r$ 
here and in Ref.~\cite{Alioli:2021wpn}, 
which should be small given their similarity.} 
missing QCD corrections by applying a constant $K$-factor of~\cite{Caola:2016trd,Alioli:2021wpn}
\begin{align}
\label{eq:kfactor}
    K^{\rm NLO} = \round{1.8594235296010675}\ . 
\end{align}

Over the range $\sqrt{s}=1-100\TeV$, the fiducial cross sections 
and relative scale uncertainties for various 
polarization configurations roughly span
\begin{subequations}
\begin{align}
    \sigma_{ZZ,\ Z_TZ_T} &\sim {10^{-2}\fb - 40\fb}\ ,
    \\
    \sigma_{Z_TZ_0} & \sim {10^{-3}\fb - 2\fb}\ ,
    \\
    \sigma_{Z_0Z_0} & \sim {10^{-3}\fb - 2\fb}\ ,
    \\
    \delta\sigma^{\rm scale}/\sigma\Big\vert_{ZZ,\ Z_\lambda Z_{\lambda'}} &\sim {\pm25\% - \pm30\%}\ .
\end{align}
\end{subequations}
We do not show rates with the auxiliary polarization $\lambda=A$ since their contraction with currents containing  massless leptons vanishes by the Dirac equation.
For lower collider energies, we observe that the $(0,0)$ rate is comparable but still lower than the mixed $(0,T)+(T,0)$ rate. At higher collider energies, however, this difference disappears.

For concreteness, the QCD-improved cross sections (first column), 
scale uncertainties [\%] (second), and 
PDF uncertainties [\%] (third)
at the LHC's current energy of $\sqrt{s}=13\TeV$ are 
\begin{subequations}
\label{eq:xsec_polar_lhc13}
\begin{align}
%
    \sigma_{ZZ} \times K^{\rm NLO}\ =\
    \pgfmathparse{1.8594235296010675*1000*0.0012273}
    \round{\pgfmathresult}&\fb\ ^{+25\%}_{-19\%}\ ^{+1\%}_{-1\%},
    \\
    \sigma_{Z_TZ_T}  \times K^{\rm NLO}\ =\
    \pgfmathparse{1.8594235296010675*1000*0.00114549}
    \round{\pgfmathresult}&\fb\ ^{+25\%}_{-19\%}\ ^{+1\%}_{-1\%},
    \\
    \sigma_{Z_0Z_T}  \times K^{\rm NLO}\phantom{1234567}\ &\ \nonumber\\
    =\
    \pgfmathparse{1.8594235296010675*10*4.61721}
    \round{\pgfmathresult}\times10^{-3}&\fb\ ^{+25\%}_{-19\%}\ ^{+1\%}_{-1\%}\ ,
    \\
    \sigma_{Z_0Z_0}  \times K^{\rm NLO}\phantom{1234567}\ \nonumber\\
    =\
    \pgfmathparse{1.8594235296010675*10*3.50175}
    \round{\pgfmathresult}\times10^{-3}&\fb\ ^{+26\%}_{-20\%}\ ^{+1\%}_{-1\%}\ ,
    \\
    \sigma_{Z_XZ_X} \times K^{\rm NLO}\ =\
    \pgfmathparse{1.8594235296010675*1000*0.00122629}
    \round{\pgfmathresult}&\fb\ ^{+25\%}_{-19\%}\ ^{+1\%}_{-1\%}\ .
\end{align}
\end{subequations}
Here and below, $Z_X$ denotes the sum over $Z_T$, $Z_0$, and $Z_A$ channels at the amplitude level and shows that the unpolarized rate is recovered.
By closure, the interference is
$\delta\sigma=\sigma_{ZZ}-\sum_{\lambda\lambda'}\sigma_{\lambda\lambda'}\sim
\mathcal{O}(+1)$ ab,
which is below our MC statistical uncertainty for 400k events.
In terms of polarization fractions
\begin{align}
    f_{\lambda\lambda}\ \equiv\ 
    \frac{\sigma_{Z_\lambda Z_\lambda}}{\sigma_{ZZ}}\ 
    =\ 
    \frac{\sigma_{Z_\lambda Z_\lambda}}{\sum_{\kappa\kappa'}\sigma_{Z_\kappa Z_{\kappa'}}}\     ,
\end{align}
the LO predictions for various channels are:
    \begin{align}
    f_{TT} = 93.3\%,& \quad 
    f_{00} = 2.9\%,
    \nonumber\\ 
    f_{T0+0T} &= 3.8\%.
    \end{align}
This is consistent with the results of Ref.~\cite{Denner:2021csi}.

In the lower panel of Fig.~\ref{fig:xsec} we show the ratio of polarized (fiducial) cross sections
$\sigma_{\lambda\lambda}$, along with their uncertainties relative to
the central unpolarized $ZZ$ rate.
The larger (transparent) bands are the
scale uncertainties at LO and reach approximately
{$\delta\sigma_{\lambda\lambda}^{\rm scale}/\sigma_{ZZ}\sim \pm20\%-\pm40\%$}.
The smaller (solid) bands are the PDF uncertainties
and reach about
{$\delta\sigma_{\lambda\lambda}^{\rm PDF}/\sigma_{ZZ}\lesssim\pm1.5\%$}
for {$\sqrt{s}\gtrsim5\TeV$}.
Specifically, we show the $Z_TZ_T$ ratio (blue bands),
which sits uniformly at around
{$(\sigma_{TT}/\sigma_{ZZ})\sim 90\%-95\%$},
and the rate of $Z_XZ_X$ (red bands) sits uniformly at unity.
The $Z_XZ_X$ ratio serves as a closure test and check of our methodology.


\subsection[\texorpdfstring{Spin-correlation in polarized $gg\to ZZ\to 4\ell$}{Spin-correlation in polarized gg to ZZ to 4l}]
{Spin-correlation in polarized $gg\to ZZ\to 4\ell$}
\label{sec:xsec_spin}

\begin{figure}[!t]
\begin{center}
\includegraphics[width=\columnwidth]{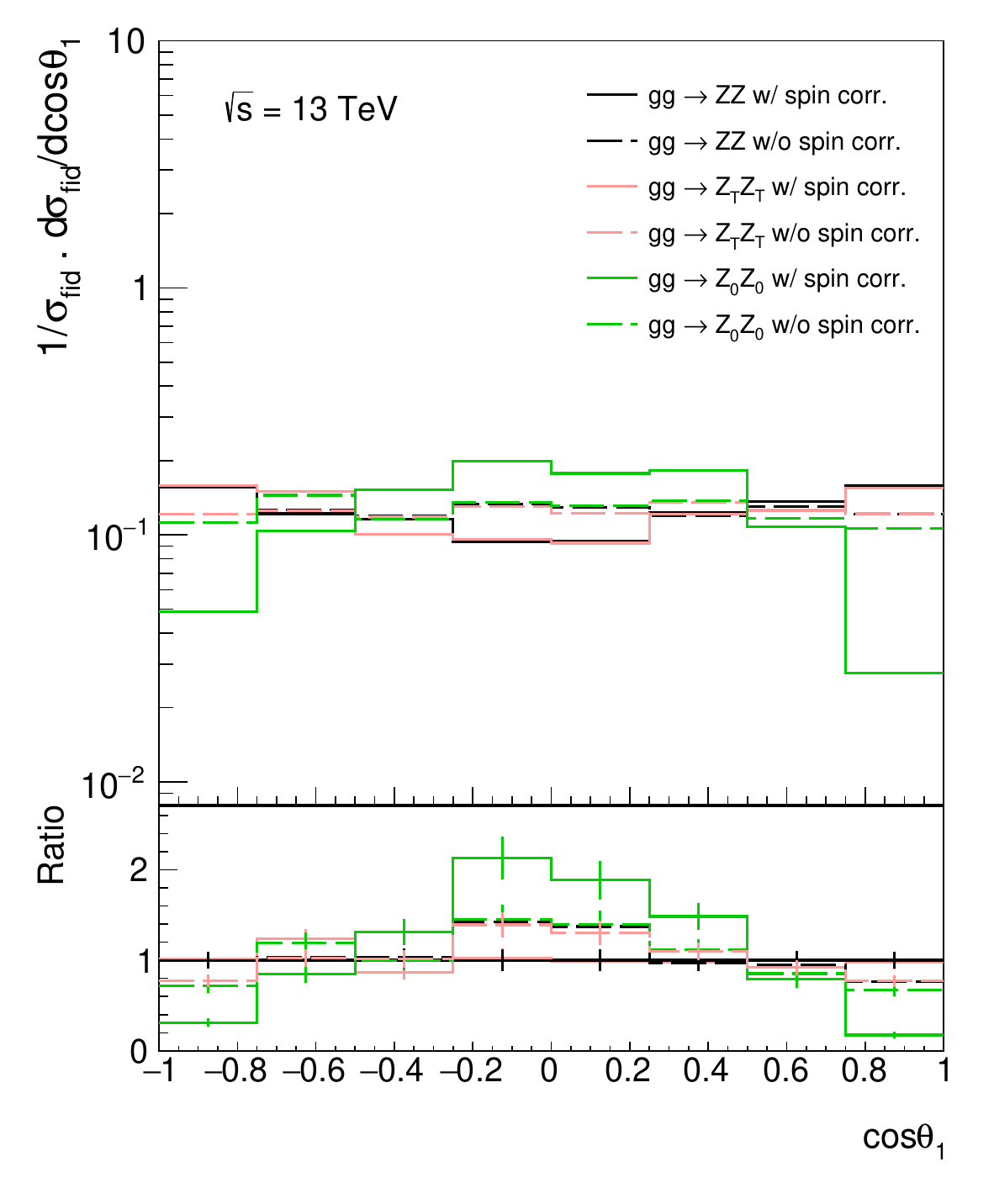}
\end{center}
	\caption{\label{fig:SpinCorr_SR} Comparison between simulated  polarized $gg\to Z_\lambda Z_{\lambda'}\to  e^+e^-\mu^+\mu^-$  events with phase space cuts of Eqs.~\eqref{eq:cuts} and \eqref{eq:cuts_tight},
 with and without spin correlation. 
 The ratios are with respect to the
 unpolarized result with spin correlation.}
\end{figure}

Measurements of $ZZ$ polarization in ggF processes require public MC generators matched to parton showers that can simultaneously model loop-induced processes, 
helicity-polarized intermediate states, 
and still preserve spin correlation.
 
Spin correlations are essential for the observables used in a typical $4\ell$-analyses. 
For example: matrix-element discriminants rely on spin correlations between final-state leptons to separate signal processes from irreducible background~\cite{ATLAS:2023dnm,ATLAS:2018jym,CMS:2022ley}.

As another demonstration of our method, 
Fig.~\ref{fig:SpinCorr_SR} shows a comparison
between simulations of the $gg\to ZZ\to 4\ell$ process with and without spin correlation. 
The variable $\theta_1$ is the angle between the three-momentum of $V_2$  and the three-momentum of the hardest fermion in the decay of $V_1$, $\ell_{11}$, 
in the rest-frame of $V_1$, 
where $V_1$ is the same-flavor $\ell^+\ell^-$ pair whose mass is closest to $M_Z$.

The $\cos\theta_1$ distribution captures the spin correlation in the process that enters the matrix-element discriminants used in Higgs analyses. A recent measurement by the ATLAS collaboration of polarization states in $q\bar{q}\to ZZ\to 4\ell$ uses a reweighting based on this variable to approximate the effect of polarization in $gg\to ZZ\to 4\ell$~\cite{ATLAS:2023zrv}. The comparison 
shows how acceptance estimates would be incorrect if spin correlations were ignored.
The largest acceptance difference is observed in $Z_0Z_0$ production (green), with smaller effects in $Z_TZ_T$ (red). 
We note the importance of spin correlation in ggF already at lowest order.



\section{Sensitivity to longitudinally polarized diboson pairs from gluon fusion}
\label{sec:PolarMeas}

We now present our projections for extracting polarization cross sections from LHC collisions.

\subsection{Template method for extracting polarization}

\begin{figure*}[!t]
\begin{center}
\subfloat[]{\includegraphics[width=0.48\textwidth]{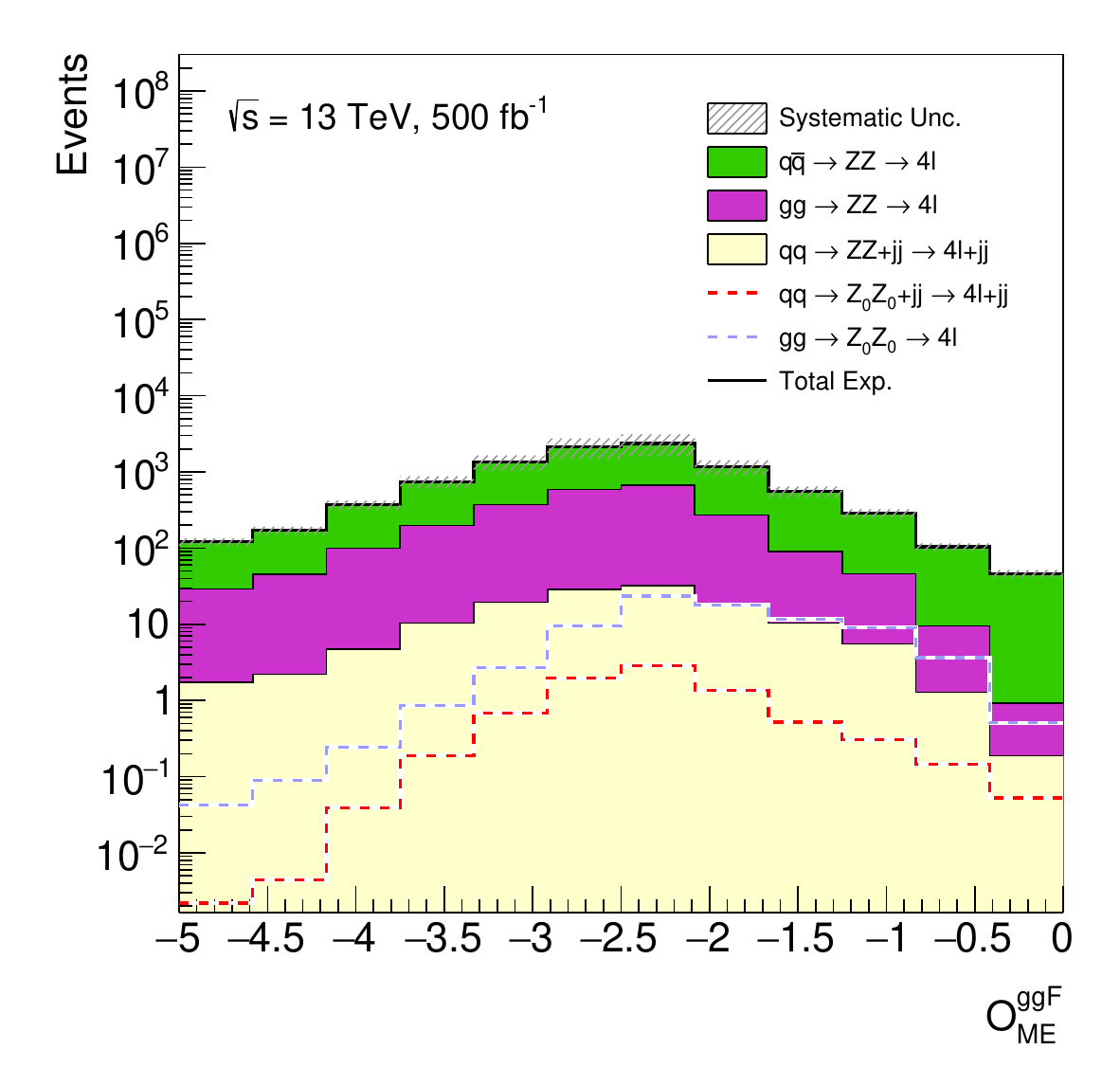}\label{fig:ggf_sr}}
\subfloat[]{\includegraphics[width=0.48\textwidth]{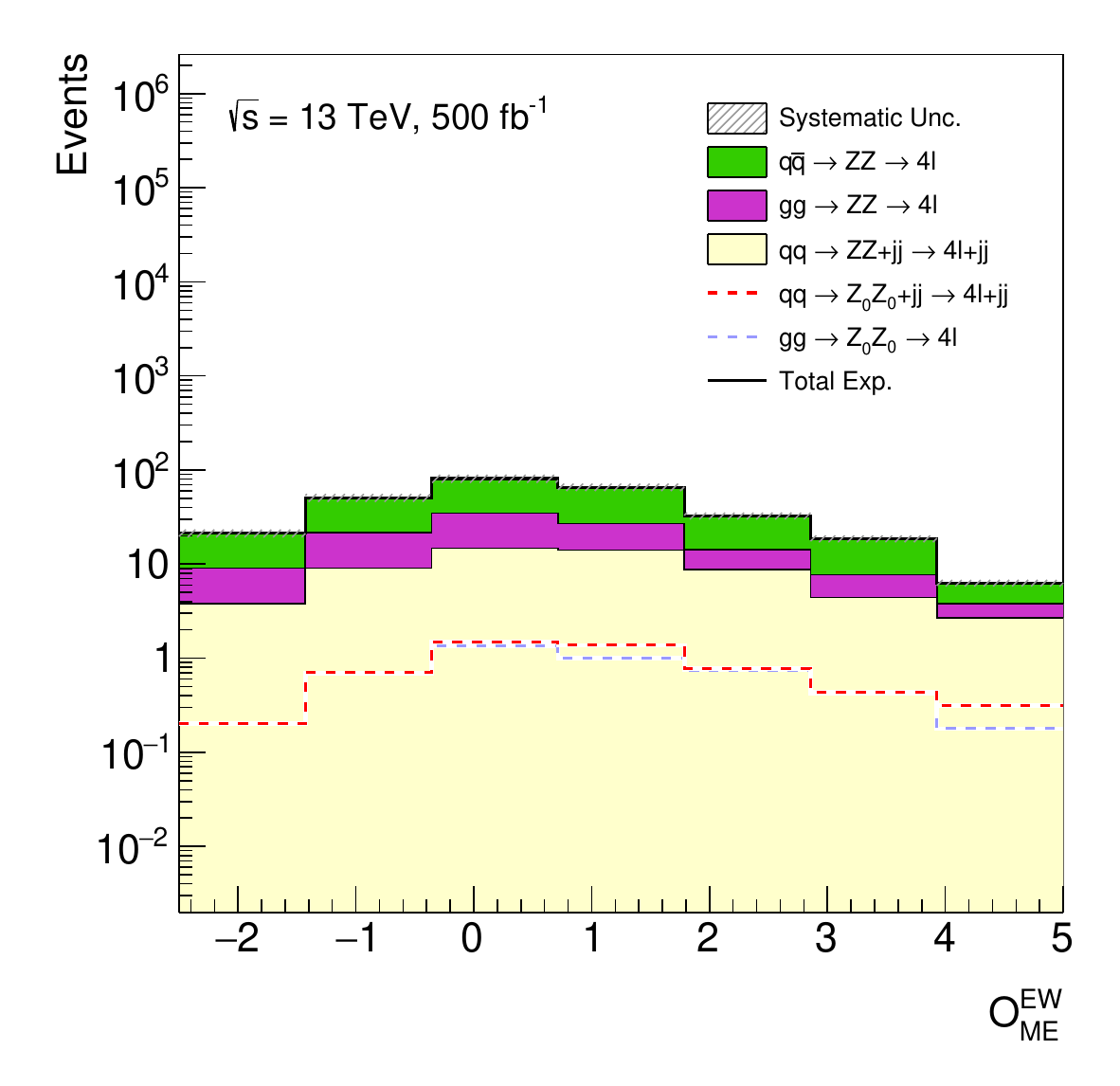}\label{fig:ew_sr}}
\end{center}
	\caption{Templates for the observables used in the (a) ggF and (b) EW regions as defined in Eq.~\eqref{eq:me_observable}. Templates are stacked and normalized to $\mathcal{L}=500\,\text{fb}^{-1}$ of integrated luminosity in $\sqrt{s}=13\TeV$ LHC collisions. Contributions from longitudinally polarized $Z_0Z_0$ pairs in ggF (blue dash) and EW (red dash) production modes are overlaid. The black solid line indicates the total expected (Total Exp.) number of events with acceptance effects from selection criteria but without taking into account detection efficiency. The hatched band is an estimate of modeling and experimental uncertainties.}
\end{figure*}

We continue our focus on  $Z_\lambda Z_{\lambda'}$ production.
Specifically, we consider two signal processes
\begin{subequations}
    \begin{align}
        \text{ggF}\ &:\ pp\ \xrightarrow{gg} Z_\lambda Z_{\lambda'} \to e^+e^-\mu^+\mu^-\ ,
        \\
        \text{EW}\ &:\ pp\ \xrightarrow{qq} Z_\lambda Z_{\lambda'} jj\to e^+e^-\mu^+\mu^- jj\ ,
    \end{align}
at $\mathcal{O}(\alpha^4\alpha_s^2)$ and $\mathcal{O}(\alpha^6)$.
The ggF channel employs the method of Sec.~\ref{sec:PolarTheory}
while the EW channel,
which includes both VBF and quark-annihilation topologies,
employs the truncated-propagator method implemented 
in Ref.~\cite{BuarqueFranzosi:2019boy}.
We also consider the (irreducible) unpolarized background at $\mathcal{O}(\alpha^4)$
\begin{align}
        \text{diboson}\ &:\ pp\ \xrightarrow{q\overline{q}} Z Z \to e^+e^-\mu^+\mu^-\ ,
    \end{align}
\end{subequations}
which we generate at NLO with \texttt{Sherpa} (v2.2)~\cite{Sherpa:2019gpd}.

We employ the so-called template method, where signal processes are treated as linear combinations of sub-processes, or templates, with unknown weights and fit these weights to (simulated) data. By using individual $Z_\lambda Z_{\lambda'}$ helicity configurations as our templates, we can identify the unknown weights as the polarization fractions $f_{\lambda\lambda'}$.

We choose the following polarization configurations as our basis of templates:
(i) $Z_0Z_0$,
(ii) $Z_TZ_T$, and
(iii) $Z_TZ_0$.
This last one is defined by subtracting the $Z_0Z_0$ and $Z_T Z_T$ channels from the unpolarized  process.
Formally, the $Z_TZ_0$ template includes a contribution from interference but this is estimated to be small by our closure test. 
Formally, the $Z_TZ_0$ template includes the $Z_A$ component and interference. For resonant diagrams, the $Z_A$ contribution is zero because we assume final-state charged leptons are massless. For simplicity, we will keep the notation $Z_TZ_0$ for all these contributions hereafter.
To maximize the sensitivity to  longitudinally polarized $Z_0Z_0$ pairs,
events are divided into two categories, 
called ``ggF'' and ``EW'', 
according to the selection criteria below.

\subsection{Event selection and observable definitions}

Simulated events are selected using criteria similar to the $H^{(*)}\to ZZ\to 4\ell$ analyses performed by ATLAS and
CMS~\cite{ATLAS:2020wny,CMS:2021ugl,CMS:2022ley,ATLAS:2023dnm}.
Specifically, we require signal and background processes to 
satisfy the selection cuts given in Eq.~\eqref{eq:cuts}
but tighten the $m_{\ell\ell}^{\rm SF}$ requirement such that
\begin{align}
\label{eq:cuts_tight}
50\GeV< m_{\ell_1\ell_2}^{\rm SF}< 106\GeV\ ,
\end{align}
where $(\ell_1\ell_2)$ is the same-flavor pair with mass closest to $M_Z$.
Jets are reconstructed by clustering all final-state hadrons and photons using the anti-k$_t$ algorithm~\cite{Cacciari:2008gp} with a radius parameter of $R=0.4$. Jet candidates are required to satisfy
\begin{align}
    p_T^j > 30\GeV\ \text{and}\ \vert\eta^j\vert < 4.4\ .
\end{align}
Jets that overlap with any signal lepton are discarded.
To increase sensitivity to the EW mode, EW events are required to additionally satisfy
\begin{align}
    n_j \geq 2,\
    \vert\Delta\eta(j_1,j_2)\vert > 3,\
    \vert\cos\theta_1\vert<0.7\ .
\end{align}
All other events are assigned to the ggF region.

For both regions, templates are built using the matrix-element (ME) kinematic discriminants 
\begin{subequations}
\label{eq:me_observable}
\begin{align}
    O_\textrm{ME}^{\text{ggF}} &= \log_{10}\nonumber
    \\
    &\ \left(\frac{|\textrm{ME}_{ggZ_0Z_0}|^2}{|\textrm{ME}_{{ggZ_TZ_T}}|^2+0.1\cdot |\textrm{ME}_{{q\overline{q}ZZ}}|^2}\right),
    \label{eq:ggF_observable}
    \\
    O_\textrm{ME}^{\text{EW}} &= \log_{10}\left(\frac{|\textrm{ME}_{qqZ_0Z_0jj}|^2}{|\textrm{ME}_{{qqZ_TZ_Tjj}}|^2}\right)\ .
    \label{eq:ew_observable}
\end{align}
\end{subequations}
The templates for the ggF and EW regions
are shown in Figs.~\ref{fig:ggf_sr} and~\ref{fig:ew_sr}, respectively.

\subsection{Statistical model and expected sensitivity}
\label{sec:stat_model}

\begin{figure*}[!t]
\begin{center}
	\subfloat[]{\includegraphics[width=0.98\columnwidth]{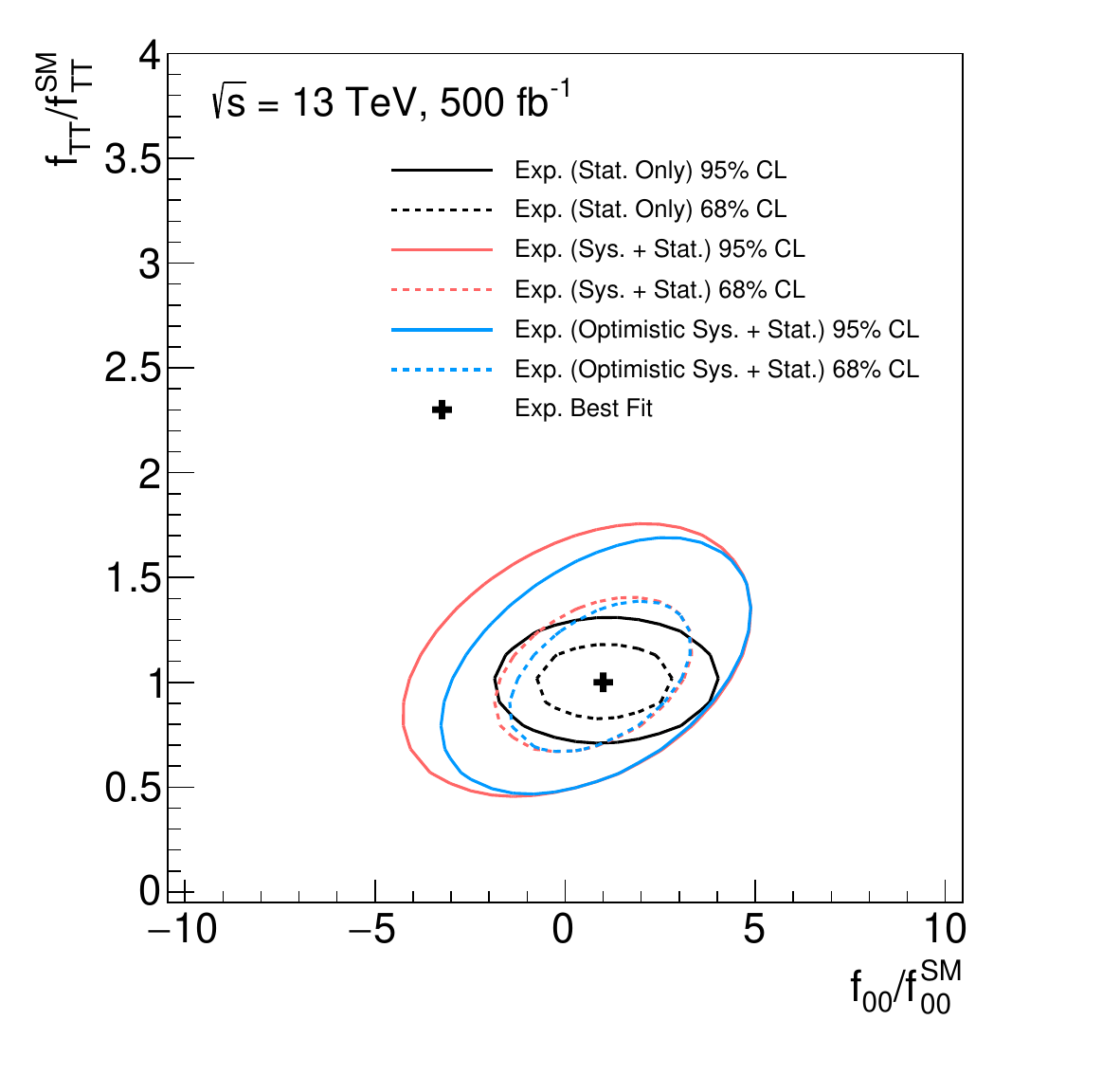}\label{FracMeas_NLLggF_Run2Run3}}
	\subfloat[]{\includegraphics[width=0.98\columnwidth]{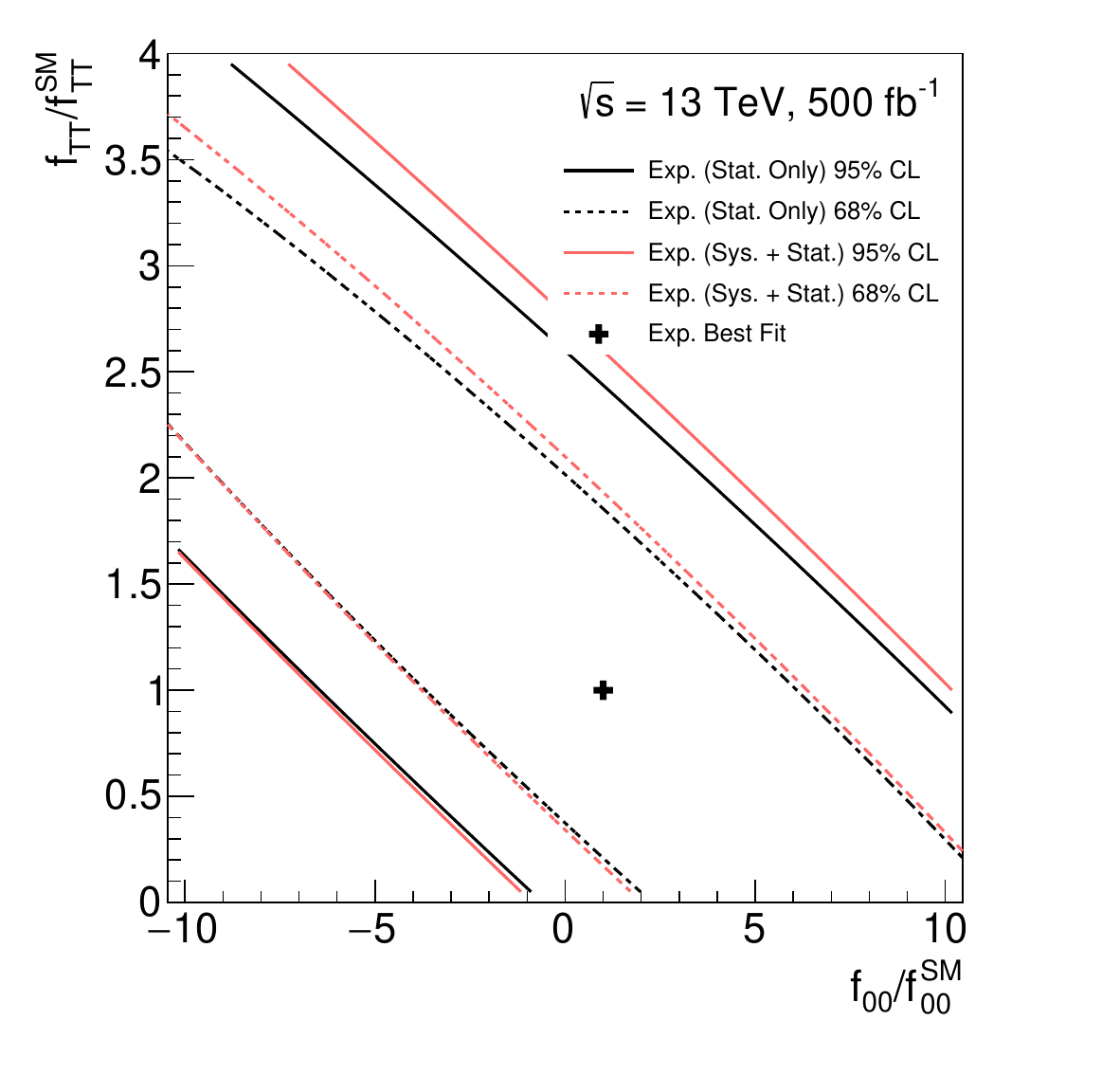}\label{FracMeas_NLLVBF_Run2Run3}}
\end{center}
	\caption{Test statistic 2D scan of helicity fractions normalized to SM predictions of $Z_\lambda Z_{\lambda'}$ in  (a) $gg\to Z_\lambda Z_{\lambda'}\to 4\ell$ and (b) EW $qq\to Z_\lambda Z_{\lambda'} jj\to 4\ell jj$,
    assuming $\mathcal{L}=500\invfb$ of LHC data.
    }
	\label{FracMeas_NLL_Run2Run3}
\end{figure*}

Expected sensitivity is assessed for two integrated luminosity scenarios at  $\sqrt{s}=13\TeV$: $\mathcal{L}=500$ and $\mathcal{L}=3000\,\text{fb}^{-1}$. The first scenario provides a good approximation for an analysis using the full Run 2 and Run 3 data sets of the LHC. The latter  corresponds to the full HL-LHC program. 

The expected sensitivity is probed by a profile log-likelihood ratio test statistic~\cite{Cowan:2010js}. The likelihood is built from the product of Poisson probability densities over the bins shown in Fig.~\ref{fig:ggf_sr} and~\ref{fig:ew_sr}, and Gaussian constraint terms to represent systematic uncertainties with nuisance parameters (NP) $\alpha$. This profile is given by
\begin{align}\label{eq:llr}
\lambda(\mu^{\text{ggF}}_{00},&\ \mu^{\text{ggF}}_{00},\alpha) = 
\nonumber\\
&\prod_{i\in \text{bins}}\text{Pois}\left[N(1,1)|N(\mu^{\text{ggF}}_{00},\mu^{\text{EW}}_{00})\right]
\nonumber\\
\times&\prod_{\alpha\in\text{NP}}\text{Gaus}(\alpha/\sigma_{\alpha}|0,1)\ ,
\end{align}
where $\sigma_{\alpha}$ is the uncertainty associated with $\alpha$.

We consider three uncorrelated sources of systematic uncertainties:
a $10\%$ experimental uncertainty on all processes,
a $25\%$ modeling uncertainty on ggF,
and a $5\%$ modeling uncertainty on all other processes. 
We also consider the optimistic scenario where ggF modeling uncertainties are reduced to $15\%$~\cite{BuarqueFranzosi:2021wrv}. 
The function $N(\mu^{\text{ggF}}_{00},\mu^{\text{EW}}_{00})$ is the  number of $ZZ\to4\ell$ events and is given by the sum of individual channels $(N^k = \mathcal{L}\times \sigma^k)$:
\begin{align}\label{eq:pol_model}
         N(\mu^{\text{ggF}}_{00},\mu^{\text{EW}}_{00})&=
         N^{q\overline{q}ZZ}
         \nonumber\\
         +N^{ggZ_TZ_T}&+N^{ggZ_TZ_0} 
         +\mu^{\text{ggF}}_{00}N^{ggZ_0Z_0}
         \\
         +N^{qqZ_TZ_Tjj}&+N^{qqZ_TZ_0jj} 
         +\mu^{\text{EW}}_{00}N^{qqZ_0Z_0jj}\ .
         \nonumber
\end{align}
Here, $\mu^{\text{ggF}}_{00}$ and $\mu^{\text{EW}}_{00}$ are signal strengths and multiply the SM predictions for $gg\to Z_0Z_0\to 4\ell$ and EW $qq\to Z_0Z_0qq\to 4\ell qq$, respectively. 
The $\mu_{00}$ are varied to maximize the log-likelihood function;
other polarization components remain unchanged.
We further take the natural logarithm of the likelihood and multiply by $-2$ ($2\text{NLL}$) to obtain a $\chi^2$ asymptotic distribution. 
The test statistic $2\text{NLL}$ is obtained by profiling the NPs and any signal strength not being inferred.

With the HL-LHC, ATLAS and CMS can constrain the production of $gg\to Z_0Z_0\to 4\ell$ ($qq\to Z_0Z_0qq\to 4\ell qq)$  to within $2.5\times$ ($4.5\times$) the SM prediction. 
The expected significance for observing $gg\to Z_0Z_0\to 4\ell$ is $1.3\sigma$ with $\mathcal{L}=3000\,\text{fb}^{-1}$, which is comparable to the expected sensitivity for $qq\to Z_0Z_0jj\to 4\ell jj$  assessed in Ref.~\cite{CMS:2018mbt}.
{The corresponding reach at Run 3 is $4\times$ ($8\times$) the SM prediction.}
By multiplying by the ratio $\sigma_{00}/\sigma_{\rm unpol.}$ with respected to the unpolarized cross-section,
 limits on $\mu_{00}$ can be translated into limits on the polarization fraction $f_{00}$.

To extract the helicity fractions of $Z_\lambda Z_{\lambda'}$ pairs $f_{\lambda\lambda'}$ (relative to the SM prediction $f_{\lambda\lambda'}^{\rm SM}$) we again use the template method.
We impose the normalization $f_{00}+f_{TT}+f_{T0+0T+\text{int.}}=1$,
where $f_{T0+0T+\text{int.}}$ accounts for both the $(0,T)+(0,T)$ polarization configurations and interference.
This reduces the number of independent helicity parameters in the fit to only two: one for the longitudinal fraction ($f_{00}/f^{\text{SM}}_{00}$) and the other for the transverse fraction ($f_{TT}/f^{\text{SM}}_{TT}$). Two statistical models are built to probe the ggF and EW channels individually.

The model for the ggF process is given by
 \begin{align}
 \label{eq:frac_model}
         &N\left(\frac{f_{00}}{f^{\text{SM}}_{00}},\frac{f_{TT}}{f^{\text{SM}}_{TT}}\right)\ 
         =\ N^{q\overline{q}ZZ}\ +\ N^{qqZZjj}\ 
         \nonumber\\
         &\ +\ 
         \left(\frac{f_{00}}{f^{\text{SM}}_{00}}\right)\cdot N^{ggZ_0Z_0}\
         +\ 
         \left(\frac{f_{TT}}{f^{\text{SM}}_{TT}}\right)\cdot N^{ggZ_TZ_T} 
         \nonumber\\
         &\ +\ \cfrac{
         s
         -\frac{f_{00}}{f^{\text{SM}}_{00}}\cdot s_{00}
         -\frac{f_{TT}}{f^{\text{SM}}_{TT}}\cdot s_{TT}}{s_{T0}}\cdot N^{ggZ_TZ_0}\ .
 \end{align}
Here, $s$, $s_{00}$, $s_{TT}$ and $s_{T0}$ are the integrals over the templates of the fiducial,  
longitudinal and transverse polarizations in both the ggF and EW regions. The expected sensitivity for measuring the values of $f_{00}/f^{\text{SM}}_{00}$ and $f_{TT}/f^{\text{SM}}_{TT}$  in ggF and EW events is shown in Figs.~\ref{FracMeas_NLLggF_Run2Run3} and \ref{FracMeas_NLLVBF_Run2Run3}, respectively.
We anticipate that projected sensitivities could be improved by employing more sophisticated analysis techniques, e.g., deep neural networks, that are common in ATLAS and CMS frameworks.


\section{Measuring Quantum Properties}
\label{sec:SpinCorr}

\begin{figure*}[!t]
\begin{center}
	\subfloat[]{\includegraphics[width=0.98\columnwidth]{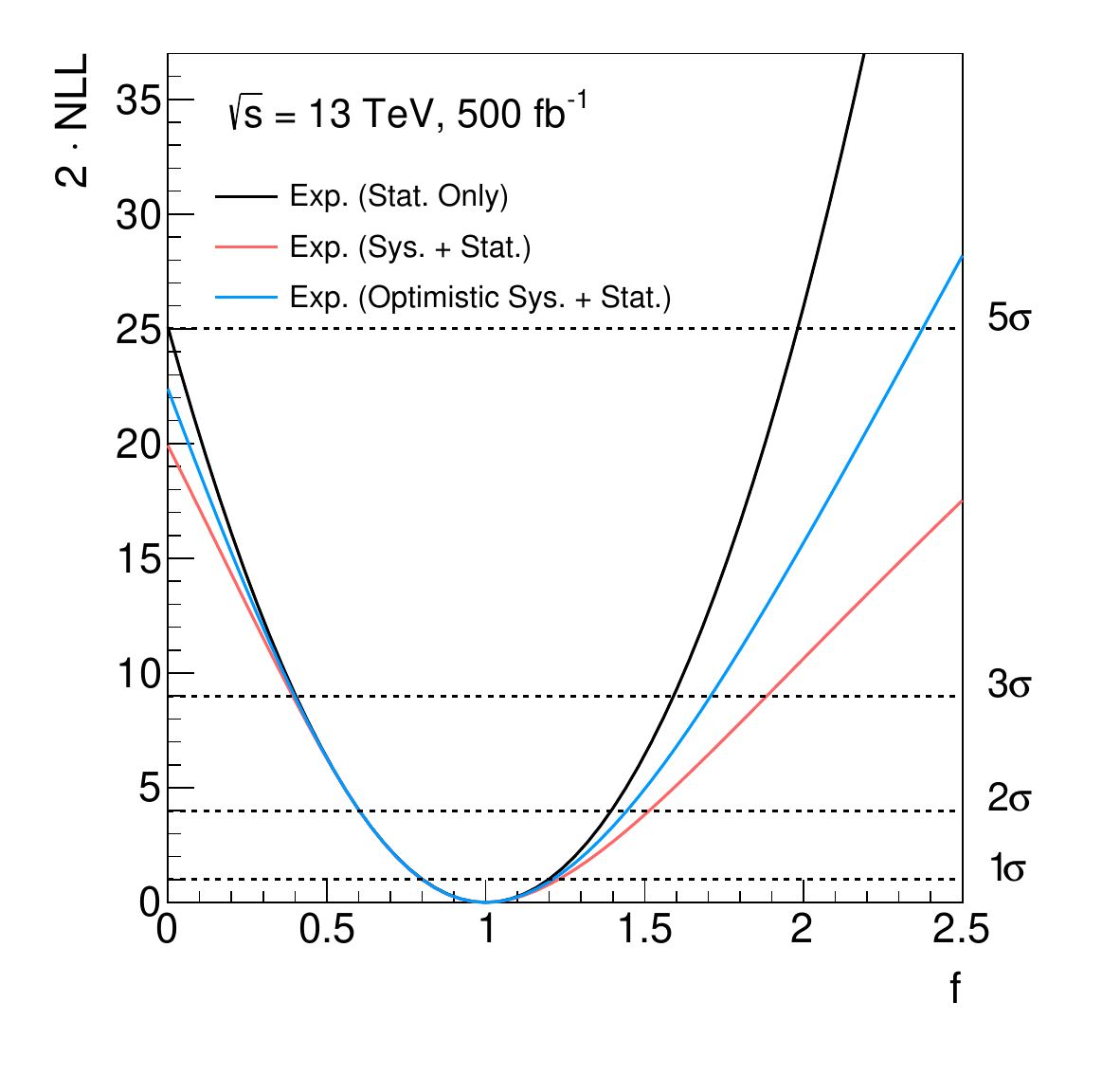}
 \label{SpinCorr_Run2Run3}}
    \subfloat[]{\includegraphics[width=0.98\columnwidth]{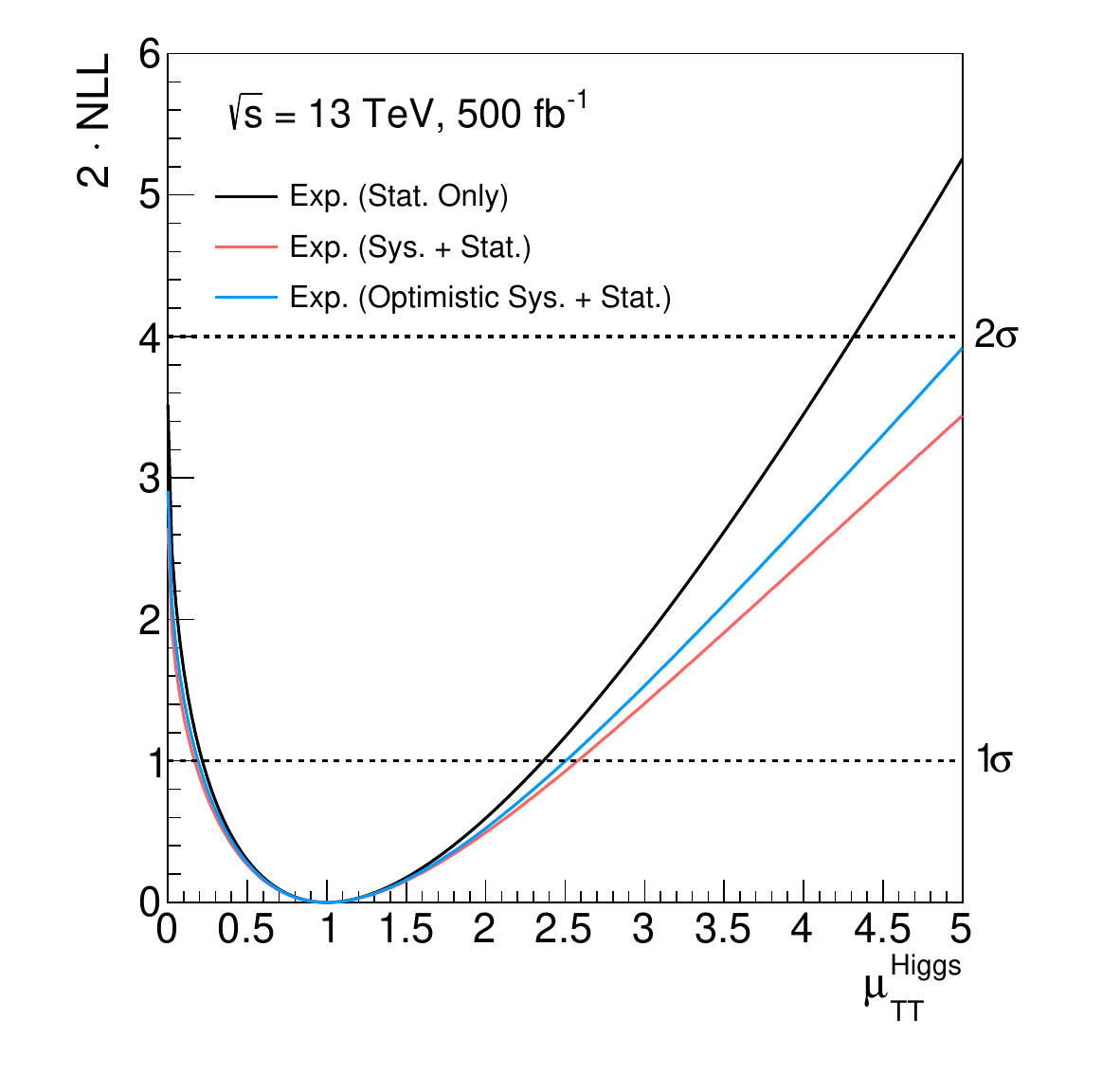}
    \label{QE_Run2Run3}}
\end{center}
	\caption{(a) Test statistics scan with respect to the parameter $f$ which represents the degree of spin-correlation in $gg\to ZZ\to 4\ell$ process,
 assuming $\mathcal{L}=500\invfb$ of LHC data.
 (b) Same as (a) but with respect to signal strength multiplying the SM production of polarized $gg\to H^*\to Z_TZ_T \to 4\ell$.
 }
\end{figure*}

Spin, and hence spin-correlation in production and decay chains of heavy particles, is a fundamental quantum property of elementary particles.
With our methodology, we can additionally propose two further tests of quantum properties in multiboson systems at high energies:
the existence of spin correlation 
and 
quantum entanglement.

To show the importance of spin correlation in $Z_{\lambda}Z_{\lambda'}$ production, we build a statistical model that assesses the sensitivity of the LHC data to spin-correlation following Ref.~\cite{ATLAS:2012ao}. 
The model 
\begin{equation}\label{eq:spin_model}
 \begin{split}
         N(f)\ & =\ N^{q\overline{q}ZZ}\ +\ N^{qqZZjj}\\
         &+\ f\cdot N_{\textrm{SM}}^{ggZZ}\
         +\ (1-f)\cdot N_{\textrm{uncorr}}^{ggZZ}\ 
 \end{split}
\end{equation}
describes the number of $ZZ\to4\ell$ events $(N)$ as a combination of the $gg\to ZZ\to 4\ell$ process  with and without spin correlation, parameterized  by $f$.

Inference on the parameter $f$, which represents the degree of observed spin correlation, is performed with the same test statistic from Eq.~\eqref{eq:llr}. Templates in $\cos\theta_1$, as shown in Fig.~\ref{fig:SpinCorr_SR}, are built for all processes. 
Fig.~\ref{SpinCorr_Run2Run3} shows that with Run 2 and 3 data alone, a $5\sigma$ exclusion of the null hypothesis (no correlation) may already be possible.

Recently, several groups have discussed the possibility of measuring quantum entanglement with $H\to VV$~\cite{Aguilar-Saavedra:2022wam,Fabbrichesi:2023cev, Ashby-Pickering:2022umy, Aguilar-Saavedra:2022mpg,Barr:2022wyq,Morales:2023gow,Aoude:2023hxv} 
and $t\bar{t}$ processes~\cite{Severi:2021cnj,Aguilar-Saavedra:2022kgy} at the LHC. 
The ATLAS collaboration also recently published the first experimental result using $t\bar{t}$ events~\cite{ATLAS:2023fsd}. 
In Ref.~\cite{Aguilar-Saavedra:2022mpg}, the authors showed that the absence of quantum entanglement in $H\to VV$ decays is equivalent to testing the hypothesis that only the $H\to V_0V_0$ polarization configuration occurs in nature.
As models testing quantum entanglement are similar to those testing polarization and spin correlation, our methodology is applicable.
 Specifically, our Feynman rules can be used to compute off-shell Higgs production with four final-state fermions when one or both intermediate weak bosons are polarized.

Here we show the LHC is able to falsify the (null) hypothesis that there is no entanglement
in $H\to VV$ splittings and hence no $gg\to H^*\to Z_TZ_T \to 4\ell$ subprocess, in the high-mass region where the Higgs boson is produced off-shell.
We extend the statistical model of Eq.~\eqref{eq:pol_model} to describe separately the contribution of diagrams with 
an $s$-channel Higgs boson (S),
the contribution from non-Higgs, ``box'' diagrams (B),
and the interference (I) between them. The method described in Sec. \ref{sec:PolarTheory} is also employed for these additional templates. The interference template is defined by subtracting the signal and background processes from the full process.
This is given by
\begin{equation}\label{eq:QE_model}
 \begin{split}
         N&(\mu^{\rm Higgs}_{TT})=N^{q\overline{q}ZZ}+N^{qqZZjj}\\
         &+N^{ggZ_0Z_0}+N^{ggZ_TZ_0}
         +N_{\rm B}^{ggZ_TZ_T}
         \\
         &+\mu^{\rm Higgs}_{TT}N_{\rm S}^{ggZ_TZ_T} + \sqrt{\mu^{\rm Higgs}_{TT}}N_{\rm I}^{ggZ_TZ_T}\ .
 \end{split}
\end{equation}
The signal strength $\mu^{\rm Higgs}_{TT}$ is the cross section for
$Z_TZ_T$ pairs normalized to the  SM prediction. 

The expected sensitivity is assessed by using templates of the observable $O_\textrm{ME}^{\text{ggF}}$ defined in Eq.~\eqref{eq:ggF_observable}. 
For Runs 2 and 3,
Fig.~\ref{QE_Run2Run3} shows that the
no entanglement hypothesis ($\mu^{\rm Higgs}_{TT}=0$)
can be rejected with approximately $2\sigma$ significance. An even stronger rejection can be obtained by 
optimizing the observable for Higgs processes.


\section{Summary and Conclusion}
\label{sec:Conclusion}

In this work we proposed a method for computing polarized cross sections that works at the level of Feynman rules.
Previous methods typically work at the amplitude or squared-amplitude levels
and require specialized event generators.
As illustrated in Eqs.~\eqref{eq:me_full}-\eqref{eq:me_decomposition},
the basic idea is to identify the sum over polarizations in a single (unpolarized) propagator subgraph 
as the sum over multiple (polarized) propagator subgraphs.
Our method's efficient implementation into existing, public simulation tools
enables realistic simulations of tree- and loop-induced processes,
including $gg\to V_\lambda V_{\lambda'}\to 4\ell,2\ell2\nu$ while maintaining spin correlation,
off-shell effects, as well as interference between resonant and non-resonant diagrams.

After exploring some phenomenological predictions for polarized $Z_\lambda Z_{\lambda'}$ pairs,
we presented sensitivity projections for measuring 
polarized $Z_\lambda Z_{\lambda'}$ pairs when produced via gluon fusion and EW processes during Run 2 and Run 3 of the LHC, and for the larger 
HL-LHC dataset.
We also presented prospects for measuring quantum properties, e.g., 
entanglement in multiboson systems.

We believe that the techniques describe in this work can be used by  experimental collaborations to further explore the production of Higgs bosons at high virtuality and 
by the theory community to explore  ramifications for new physics. 
While we have focused only on a SM case study, 
the basic idea can be applied to resonances in new physics scenarios
as well as to weak bosons and the top quark in the
Standard Model Effective Field Theory.
Finally, there are field-theoretic aspects of ``polarization as a Feynman rule'' that merit further investigation and we encourage such studies.


\section*{Acknowledgements}

RR thanks the CalTech theory group for hosting him while this work was completed. The authors also thank Benjamin Fuks, Verena Martinez, Olivier Windu Mattelaer, and Giovanni Pelliccioli for useful discussions.

The work of MJ, RCLSA, and JS is supported in part by the DOE grant DE-SC0010004.
RR acknowledges the support of the Narodowe Centrum Nauki under Grants No.\ 2019/34/E/ST2/ 00186 and
No.\ 2023/49/B/ST2/04330 (SNAIL). The author also acknowledges the support of the Polska Akademia Nauk (grant agreement PAN.BFD.S.BDN.613.022.2021 - PASIFIC 1, POPSICLE).
This work has received funding from the European Union's Horizon 2020 research and innovation program under the Sk{\l}odowska-Curie grant agreement No.  847639, and from the Polish Ministry of Education and Science.
The authors would like to acknowledge the contribution of the COST Action CA22130 (COMETA).


\appendix

\section{Implementation details}
\label{app:implementation}

For the transverse and longitudinal polarization vectors, 
we adopt the \texttt{HELAS} convention~\cite{Murayama:1992gi}.
For the auxiliary polarization vector, we follow
Ref.~\cite{BuarqueFranzosi:2019boy} and use
\begin{align}
    \varepsilon^\mu(q,\lambda=A) = \frac{q^\mu}{M_V}\sqrt{\frac{q^2-M_V^2}{M_V^2}} \ .
    \label{eq:polar_aux_app}
\end{align}

To implement the methodology described in Sec.~\ref{sec:PolarTheory},
we start from the full \texttt{FeynRules}
SM model file 
\texttt{sm.fr}~\cite{Christensen:2008py,Christensen:2009jx,Degrande:2011ua,Alloul:2013bka}
and introduce the particles
\begin{subequations}
\begin{align}
Z   &: \texttt{ZX, Z0, ZT, ZA}\ ,\\
W^+ &: \texttt{WX+, W0+, WT+, WA+}\ ,\\
W^- &: \texttt{WX-, W0-, WT-, WA-}\ ,
\end{align}
\end{subequations}
by making several copies of \texttt{V[n]} objects.
We ensure that the appropriate values and classifications for internal/external parameters are inherited.
We then make the field (re)definition 
\begin{verbatim}
...
Definitions ->
    {Z[mu_]->Z0[mu]+ZT[mu]+ZA[mu]+ZX[mu]}
...
Definitions -> 
    {W[mu_]->W0[mu]+WT[mu]+WA[mu]+WX[mu]}
\end{verbatim}
and remove new but unphysical two-point vertices.
We also (re)assign the following PIDs:
\begin{subequations}
\begin{align}
Z(239),\ &\ W^\pm(\pm249)\ , \\
Z_X(23),\ &\ W_X^\pm(\pm24)\ , \\
Z_0(230),\ &\ W_0^\pm(\pm240)\ ,  \\
Z_T(231),\ &\ W_T^\pm(\pm241)\ ,  \\
Z_A(232),\ &\ W_A^\pm(\pm242)\ .  
\end{align}
\end{subequations}
The fields $Z(239)$ and $W^\pm(\pm249)$ are designated \texttt{Unphysical->True}, meaning that they do not appear in the final UFO library. 
Importantly, a feature in {\mgamc} is to override particles labels when PIDs match those of SM particles.
This means that in {\mgamc}, the unpolarized fields 
$Z_X(23)$ and $W_X^\pm(\pm24)$ are \textit{automatically} relabeled 
\texttt{z}, \texttt{w+/w-}.

We extract QCD renormalization and $R_2$ counter terms up to the first order in $\alpha_s$ using 
\texttt{FeynRules} {(v2.3.36)} with
\texttt{NLOCT} {(1.02)}~\cite{Degrande:2014vpa}
and \texttt{FeynArts} {(v3.11)}~\cite{Hahn:2000kx},
and package them into the {\libName} UFO,
which is publicly available from \href{https://feynrules.irmp.ucl.ac.be/wiki/VPolarization}{feynrules.irmp.ucl.ac.be/wiki/VPolarization}.
Finally, we modify (by hand) the \texttt{particles.py}
file so that the ``new'' particles have propagators given by
\begin{subequations}\label{eq:part_def}
\begin{align}
\texttt{VX}   &: \Pi_{\mu\nu}^\texttt{VX} = \sum_{\lambda\in\{\pm1,0,A\}} \Pi_{\mu\nu}^{V\lambda}\ ,\\
\texttt{V0} &: \Pi_{\mu\nu}^\texttt{V0} = \sum_{\lambda\in\{\pm1\}} \Pi_{\mu\nu}^{V\lambda}\ ,\\
\texttt{VT} &: \Pi_{\mu\nu}^\texttt{VT} = \Pi_{\mu\nu}^{V\lambda=0}\ ,\\
\texttt{VA} &: \Pi_{\mu\nu}^\texttt{VA} = \Pi_{\mu\nu}^{V\lambda=A}\ ,
\end{align}
\end{subequations}
where $\Pi_{\mu\nu}^{V_\lambda}$ is the truncated propagator   in Eq.~\eqref{eq:trunc_prop}.

\section{Test statistic scan for polarization extraction}
\label{app:stats}

\begin{figure*}[!t]
\begin{center}
	\subfloat[]{\includegraphics[width=0.97\columnwidth]{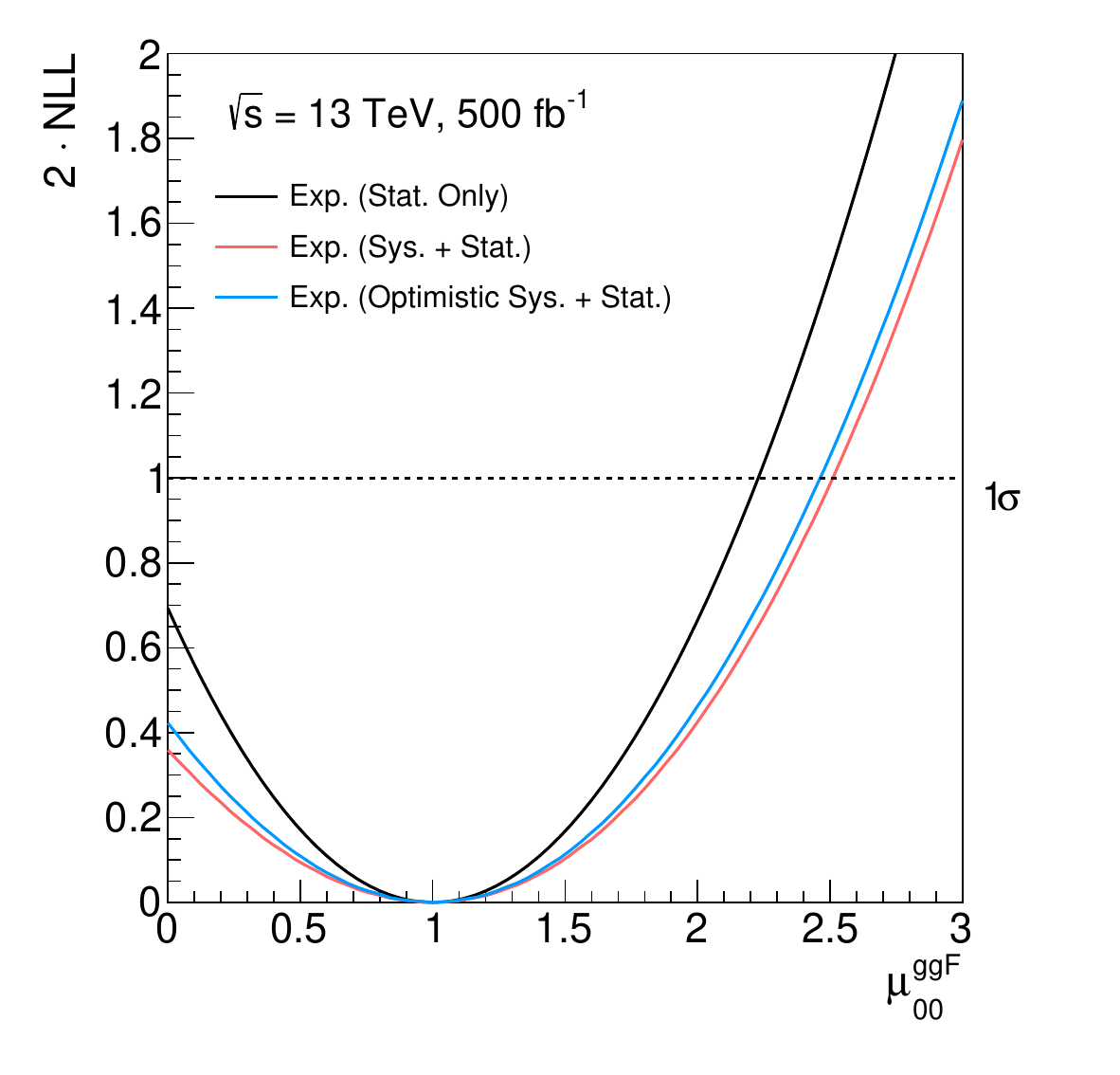}\label{PolarMeas_NLLggF_Run2Run3}}
	\subfloat[]{\includegraphics[width=0.97\columnwidth]{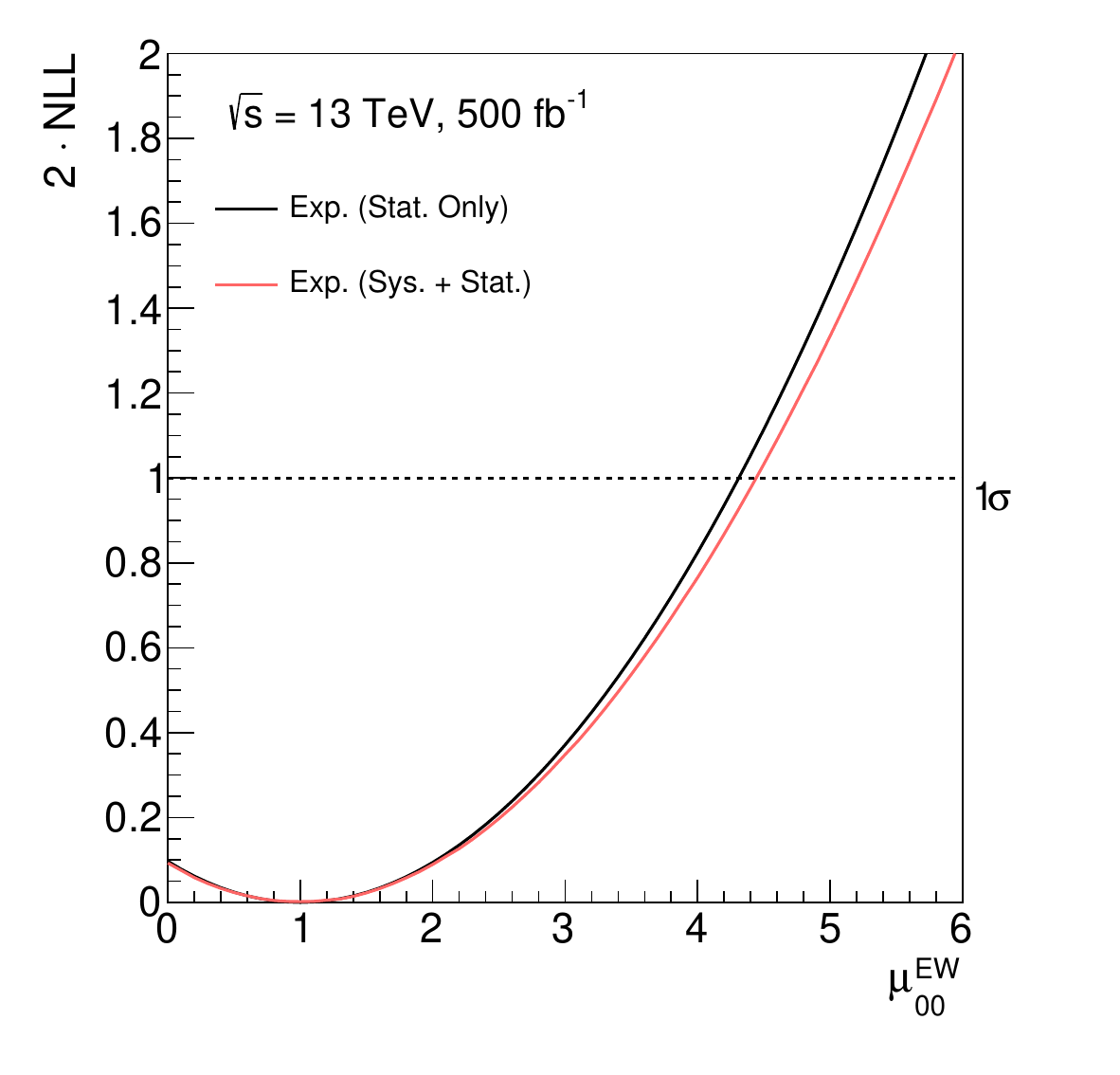}\label{PolarMeas_NLLVBF_Run2Run3}}
 \\
 	\subfloat[]{\includegraphics[width=0.97\columnwidth]{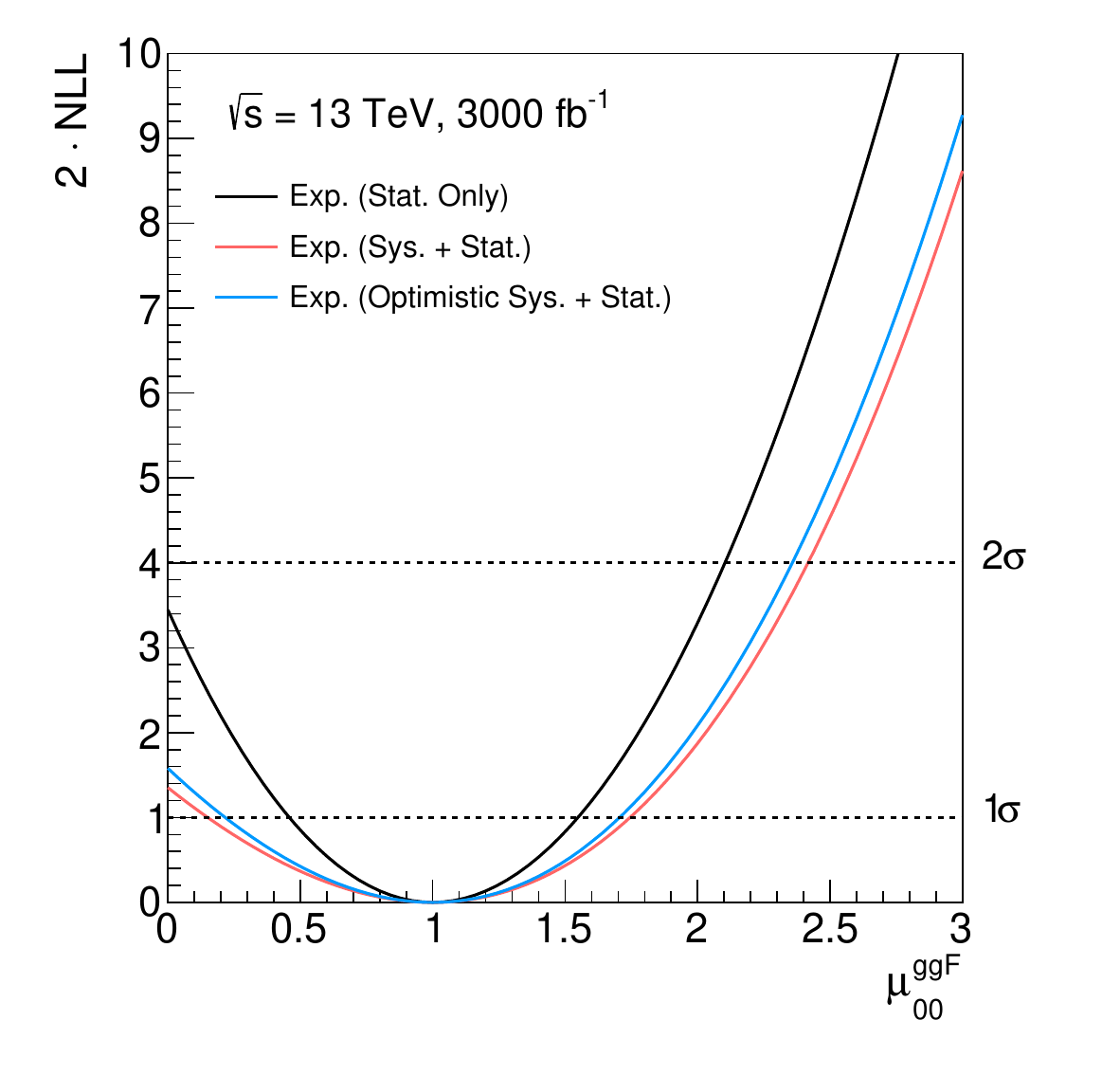}\label{PolarMeas_NLLggF_HLLHC}}
	\subfloat[]{\includegraphics[width=0.97\columnwidth]{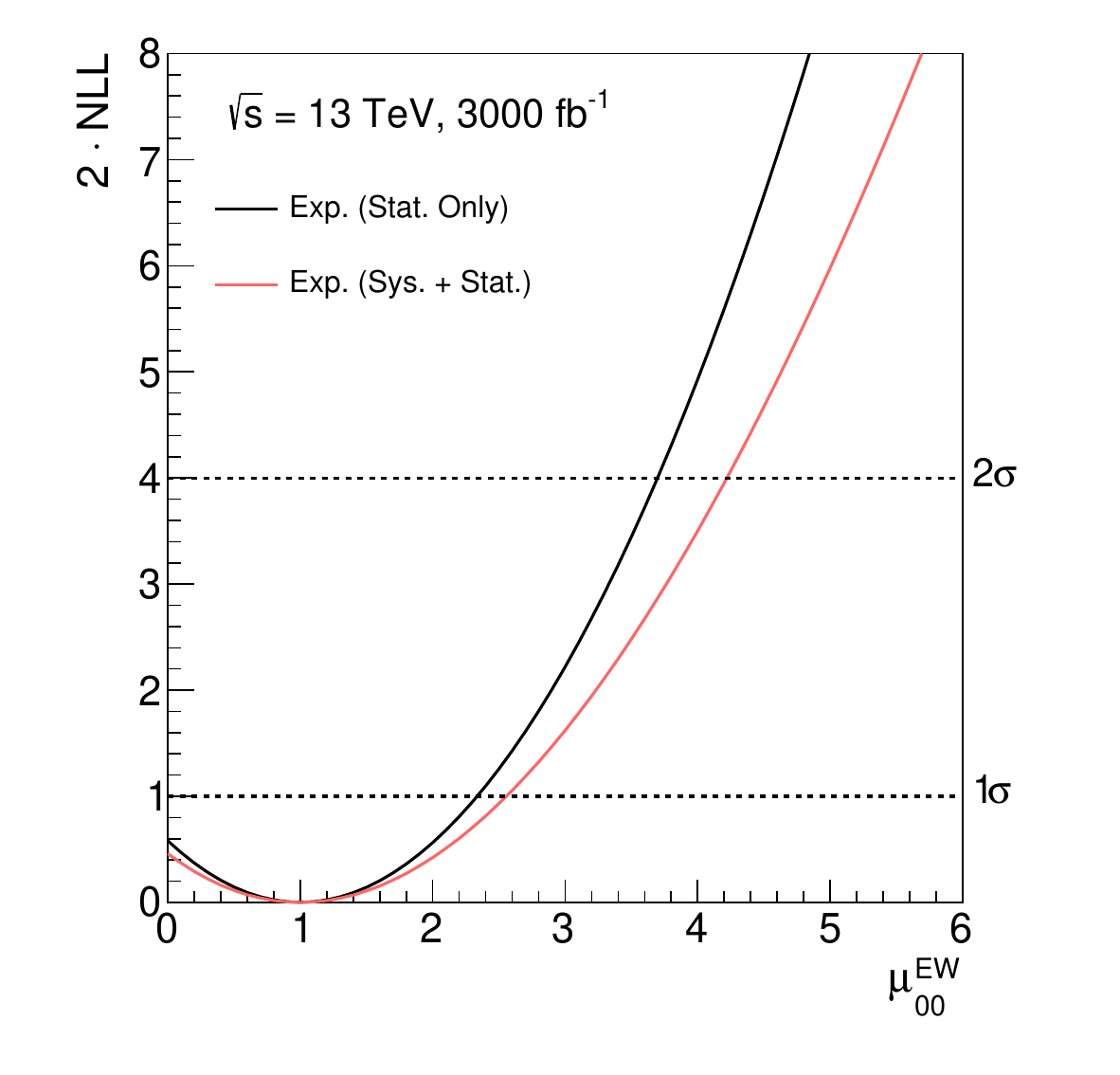}\label{PolarMeas_NLLVBF_HLLHC}}
\end{center}
	\caption{Test statistic scan with respect to signal strength multiplying the SM production of polarized (a) $gg\to Z_0Z_0\to 4\ell$ and (b) EW $qq\to Z_0Z_0jj\to 4\ell jj$ assuming $\mathcal{L}=500\invfb$ of LHC data at $\sqrt{s}=13\TeV$.
 (c,d) Same as (a,b) but for $\mathcal{L}=3000\invfb$.
 }
	\label{fig:PolarMeas_NLL}
\end{figure*}

Figure~\ref{fig:PolarMeas_NLL}  shows the expected sensitivity for the observation of longitudinally polarized $Z_0Z_0$ pairs in (a,c) $gg\to ZZ\to 4\ell$ and (b,d) $qq\to ZZjj\to 4\ell jj$ production in the high-mass regime
(a,b) after Runs 2 and 3 of LHC data taking, i.e., $\mathcal{L}=500\invfb$ at $\sqrt{s}=13\TeV$,
as well as 
(c,d) with the full HL-LHC data set, i.e., $\mathcal{L}=3000\invfb$.

\section{Computational Setup}
\label{app:setup_mc}

We assume $n_f=5$ massless quarks, with a Cabibbo-Kobayashi-Maskawa matrix equal to the identity matrix, i.e., no quark flavor mixing. Other SM inputs are set to the following values:
\begin{align}
m_t(m_t) &= 173.3\GeV,    m_h = 125.7\GeV,  \nonumber\\
M_Z 	&= 91.1876\GeV,   \alpha_{\rm QED}^{-1}(M_Z)=127.94, \nonumber\\
G_F     &= 1.174560\cdot 10^{-5}\GeV^{-2}. \label{eq:sminputs}
\end{align}

The central $(\zeta=1)$ collinear factorization and renormalization  scales are set to be half the sum of the transverse energy of final-state particles $(f)$:
\begin{align}
    \mu_f,\ \mu_r = \zeta \times \mu_0,\
     \mu_0 \equiv \sum_f \sqrt{m_f^2 + p_{Tf}^2}\ .
\end{align}
Scale uncertainties are estimated by varying $\zeta$ discretely over the range $\{0.5,1.0,2.0\}$.
For select processes, we include up to one additional parton in the matrix element using the multi-leg matching (MLM)~\cite{Mangano:2002ea,Mangano:2006rw}.
For MLM in the gluon fusion channel, diagrams in which neither a $Z$ nor a $H$ is attached to the quark loop are filtered out since they are 
part of the $q\overline{q}$ annihilation channel.
Simulated events are showered and hadronized using \texttt{Pythia8}~\cite{Sjostrand:2014zea} {(\texttt{tune=pp:14})} and matched to the
NNPDF23qed LO PDF {(\texttt{PDF:pSet=13})} \cite{Ball:2013hta}.
Hadrons are clustered using \texttt{FastJet}~\cite{Cacciari:2005hq,Cacciari:2011ma}.

Using {\libName} in {\mgFull} (\mgamc), loop-induced processes such as Eq.~\eqref{eq:proc_gf}
can be simulated at lowest order and be passed to other programs for further processes, e.g., parton showering.
For $gg \to Z_{T}Z_{T} \to e^+e^-\mu^+\mu^-$, 
the relevant {\mgamc} syntax to build matrix elements 
that includes spin-correlation, off-shell effects, and interference
is:
\begin{verbatim}
import model SM_Loop_ZPolar
generate g g > e+ e- mu+ mu- QED=4 QCD=2
        [noborn = QCD] / a z z0 za
output ZPolar_gg_eemm_TT_QED4_QCD2_XLO
\end{verbatim}

The \texttt{g g > e+ e- mu+ mu-} syntax will 
interfere all instances of $Z_\lambda Z_{\lambda'}$.
To filter out unwanted diagrams, including photon contributions,
the ``\texttt{/ a z z0 za}'' syntax is added.
The matrix elements for the analogous (i) $Z_0Z_0$, (ii) $Z_AZ_A$, and (iii) $Z_XZ_X$ channels can be generated using the filters:
(i) \texttt{/a z za zt},
(ii) \texttt{/ a z z0 zt},
(iii) \texttt{/ a z}.
For the mixed-polarization configuration
$Z_0Z_T+Z_TZ_0$, we apply the diagram selection
described in Sec.~\ref{sec:PolarTheory_implement}
and available from the repository
given in Footnote~\ref{foot:gitlab}.
The relevant {\mgamc} syntax is then
\begin{verbatim}
generate    g g > e+ e- mu+ mu- QED=4 QCD=2
        [noborn = QCD] /
        a --loop_filter=True
\end{verbatim}

The cross sections in Eq.~\eqref{eq:xsec_polar_lhc13}
  can be obtained using the following (and similar)
  commands:
\begin{verbatim}
launch ZPolar_gg_eemm_00_QED4_QCD2_XLO
analysis=off
set pdlabel lhapdf
set lhaid 324900
set lhc 13
set nevents 4k
set dynamical_scale_choice 3
set me_frame [3,4,5,6]
set use_syst true
set no_parton_cut
set mmll 50
set mmllmax 115
set mmnl 220
set etal 2.7
set drll 0.1
set ptl1min 20
set ptl2min 15
set ptl3min 10
set ptl4min 5
done
\end{verbatim}


{\small
\bibliographystyle{elsarticle-num}
\biboptions{sort&compress}
\bibliography{ggZZPolar_refs.bib}
}
\end{document}